 \documentclass[preprints,review,accept,pdftex,moreauthors]{Definitions/mdpi}

\firstpage{1}
\pubvolume{1}
\issuenum{1}
\articlenumber{0}
\pubyear{2026}
\copyrightyear{2026}
\datereceived{}
\daterevised{}
\dateaccepted{}
\datepublished{}

\usepackage{bm}
\usepackage{mathtools}
\usepackage{slashed}
\usepackage{dsfont}                         

\usepackage{tikz}                           
\usepackage[compat=1.1.0]{tikz-feynman}     
\tikzfeynmanset{warn luatex=false}          
\newcommand{\iu}{{i\mkern1mu}}

\newcommand{\lineseparator}[3]{%
    \begin{center}\begin{tikzpicture}
        \draw[{#2}-{#3}, semithick, {#1}] (0,0) to (0.65\linewidth,0);
    \end{tikzpicture}\end{center}%
}

\theoremstyle{definition}
\newtheorem{unformattedtheorem}{Theorem}[section]

\theoremstyle{definition}
\newtheorem{unformatteddefinition}{Definition}[section]

\theoremstyle{remark}
\newtheorem{unformattedremark}{Remark}[section]

\theoremstyle{definition}
\newtheorem{unformattedcorollary}{Corollary}[section]

\theoremstyle{remark}

\newcommand{\ket}[1]{\big\vert #1\big\rangle}
\newcommand{\bra}[1]{\big\langle #1\big\vert}
\newcommand{\overlap}[2]{\big\langle #1\big\vert #2\big\rangle}

\newcommand{\planewaveket}[2]{\big\vert #1;#2\big\rangle}
\newcommand{\planewavebra}[2]{\big\langle #1;#2\big\vert}

\DeclareMathOperator{\imaginary}{Im}
\DeclareMathOperator{\real}{Re}

\DeclareMathOperator*{\sumintegral}{%
\mathchoice
{\ooalign{$\displaystyle\sum$\cr\hidewidth$\displaystyle\int$\hidewidth\cr}}
{\ooalign{\raisebox{.14\height}{\scalebox{.7}{$\textstyle\sum$}}\cr\hidewidth$\textstyle\int$\hidewidth\cr}}
{\ooalign{\raisebox{.2\height}{\scalebox{.6}{$\scriptstyle\sum$}}\cr$\scriptstyle\int$\cr}}
{\ooalign{\raisebox{.2\height}{\scalebox{.6}{$\scriptstyle\sum$}}\cr$\scriptstyle\int$\cr}}
}

\newcommand{\carbonfermidecay}{{}^{10}\mathrm{C}\rightarrow{}^{10}\mathrm{B}}
\newcommand{\oxygenfermidecay}{{}^{14}\mathrm{O}\rightarrow{}^{14}\mathrm{N}}

\Title{Recent Progress in Ab-Initio Nuclear Theory for Precision Physics Searches in Muonic Atoms and Superallowed \texorpdfstring{$\beta$ Decays}{Beta Decays}}

\Author{Simone Salvatore Li Muli $^{1,}$* and Michael Gennari $^{2}$}

\AuthorNames{Simone Salvatore Li Muli and Michael Gennari}

\address{%
$^{1}$ \quad Department of Physics, Chalmers University of Technology, SE-412 96 Gothenburg, Sweden; simone.limuli@chalmers.se\\
$^{2}$ \quad Institut f{\"u}r Kernphysik and PRISMA\textsuperscript{++} Cluster of Excellence, Johannes Gutenberg-Universit{\"a}t Mainz, 55128 Mainz, Germany; michael.gennari@uni-mainz.de}

\corres{Correspondence: simone.limuli@chalmers.se}

\abstract{Precision tests of the Standard Model at low energy are increasingly limited by nuclear-structure theory rather than by experiment. We review two such cases: the two-photon-exchange correction to the Lamb shift in muonic atoms, and the \texorpdfstring{$\gamma W$}{gamma-W} box radiative correction to superallowed \texorpdfstring{$\beta$}{beta} decays. Although they probe different physics, both are governed by the same generalized hadronic tensor, so that the chiral effective field theory Hamiltonians and currents, Lanczos-based response methods, and Bayesian uncertainty quantification developed for one carry over directly to the other. We summarize recent ab initio progress in light nuclei and its impact on nuclear charge radii, on the helium isotope-shift puzzle, and on the extraction of \texorpdfstring{$V_{ud}$}{Vud} for the top-row CKM unitarity test, and state a future outlook.}


\begin{document}

\section{Introduction}
\label{sec:introduction}

The Standard Model (SM) of particle physics, despite its extraordinary empirical success, leaves fundamental questions unanswered. For example, the origin of neutrino masses, the nature of dark matter, and the baryon asymmetry of the universe, among others. It is therefore widely regarded as the low-energy limit of a more complete theory. Searches for its extensions proceed along two complementary lines. At the \emph{energy frontier}, collider experiments aim to produce new particles directly. At the \emph{precision frontier}, one confronts observables that can be both measured and computed with high accuracy where virtual contributions of new interactions would appear as small but systematic deviations between experiment and SM prediction~\cite{arXiv:1907.02164, Bro23}. In an effective-field-theory (EFT) description, new interactions arising at a scale $\Lambda$ induce low-energy couplings suppressed by $(v/\Lambda)^2$, with $v$ the electroweak vacuum expectation value, so that experiments sensitive to relative deviations at the $ \mathcal{O}(10^{-3})$ to $\mathcal{O}(10^{-4})$ level probe effective scales of order $10$~TeV and beyond, that is to say, at or above the direct reach of the Large Hadron Collider~\cite{arXiv:1907.02164, Bro23}. Atomic nuclei and atomic bound states, which can be prepared and measured with exceptional control, host several flagship tests of this kind~\cite{Saf18}; this review is concerned with two of them, and with the nuclear theory that underpins their interpretation.

Nuclear $\beta$ decay was instrumental in establishing the $V\!-\!A$ structure of the weak interaction and remains today one of the most sensitive low-energy probes of the charged-current sector~\cite{arXiv:1907.02164, Bro23}. Superallowed $0^+\to0^+$ Fermi transitions provide the most precise determination of the up-down quark mixing matrix element $V_{ud}$ in the Cabibbo-Kobayashi-Maskawa (CKM) matrix, the quantity which dominates the top-row unitarity test $ \Delta_{\mathrm{CKM}} = |V_{ud}|^2 + |V_{us}|^2 + |V_{ub}|^2 - 1 $~\cite{PRC.102.045501.2020, ARNPS.74.2347.2024}. When dispersive evaluations of the single-nucleon electroweak radiative corrections reduced their uncertainty by a factor of two~\cite{PRL.121.241804.2018}, the unitarity measure $\Delta_{\mathrm{CKM}}$ moved into tension with the SM. At present, the top-row sum falls short of unity at the level of two to three standard deviations (depending on the treatment of the experimental input) and is known as the Cabibbo angle anomaly~\cite{Nav24, ARNPS.74.2347.2024}. Beyond the unitarity test, precision measurements of $\beta$-spectra and decay correlations probe scalar and tensor interactions that are absent in the SM: per-mille-level bounds on the Fierz interference term translate into sensitivity to new physics above the $10$~TeV scale. This is competitive with and complementary to direct collider searches~\cite{arXiv:1907.02164, Bro23,Gli22,Kin23}.

Precision spectroscopy of light atomic systems offers a second, independent test. Hydrogen-like atoms are among the most well-understood bound states in nature theoretically speaking. Comparisons of their measured and computed level structures tests quantum electrodynamics (QED) for composite systems, determines fundamental parameters such as the Rydberg constant, and constrains hypothetical new lepton-nucleus interactions at the femtometer scale~\cite{Pac23}. Since the muon is roughly $200$ times heavier than the electron, muonic atoms play a special role in this program; nuclear finite-size effects in the spectra are enhanced by the third power of the mass ratio, which makes the muonic Lamb shift the most precise source of absolute nuclear charge radii~\cite{Ant22a}. The charge radius extracted from muonic hydrogen famously disagreed with the electronic determinations accepted at the time -- known as the proton-radius puzzle~\cite{Poh10,Ant13}, a discrepancy since resolved in favor of the muonic value -- and muonic radii now enter the CODATA adjustment of fundamental constants~\cite{Tie21}. Confronting muonic with electronic determinations has since developed into a sensitive test of lepton universality, with recent focus having fallen on the helium-isotope shift. The muonic and electronic values for the difference of the squared charge radii of $^3$He and $^4$He were found to disagree at the $3.6\sigma$ level~\cite{Kra21,Sch25,Van23}, and a systematic re-evaluation of the nuclear-structure corrections sharpened the tension to $4\sigma$~\cite{Lim25}. The discrepancy has since been traced to atomic rather than nuclear theory. Known second-order hyperfine contributions to the electronic isotope shift, previously neglected or treated only approximately, were recently identified and computed~\cite{Qi25,Pac24}, which brought the two determinations into agreement at the $1.3\sigma$ level with respect to Ref.~\cite{Sch25}, and $1.7\sigma$ with respect to Ref.~\cite{Lim25}. This turned the helium isotope shift from a puzzle into a stringent consistency test of the muonic-atom program.

In all of these searches, the sensitivity to new physics is limited not by experiment but by nuclear structure theory. In muonic atoms, the QED contributions are under excellent control \cite{Rat26}, and the uncertainty of the extracted radii is dominated by the two-photon-exchange (TPE) or $ \gamma \gamma $ box correction in which the exchanged virtual photons are sensitive to the full nuclear spectrum~\cite{Ji18,Pac23}. In superallowed decays, the nuclear-structure-dependent radiative corrections, particularly the nuclear $\gamma W$ box, were found to be far less controlled than previously assumed once the single-nucleon analysis had been improved. Their re-evaluation inflated the uncertainty in $V_{ud}$ by roughly $50\%$ between the 2015 and 2020 critical surveys~\cite{PRC.91.025501.2015, PRL.123.042503.2019, PRC.102.045501.2020, Bro23}. Likewise, interpreting spectrum-shape and correlation measurements at their targeted $ \mathcal{O}(10^{-3})$ accuracy requires recoil-order and induced-current nuclear corrections of commensurate precision~\cite{arXiv:1907.02164, Gli22,Kin23}. This entire program therefore hinges on a single capability: computing amplitudes of electroweak currents in nuclei at the percent level with rigorous uncertainties that admit a statistical interpretation. This is what modern \textit{ab initio} nuclear theory has begun to deliver by combining nuclear Hamiltonians and electroweak currents consistently derived from chiral effective field theory ($\chi$EFT)~\cite{Epe09,Mac11,Ham20}, systematically improvable many-body methods~\cite{Her20}, and Bayesian uncertainty quantification~\cite{Fur15,Mel19}.

\begin{figure}[t!]
    \begin{center}
    \includegraphics[width=0.75\linewidth]{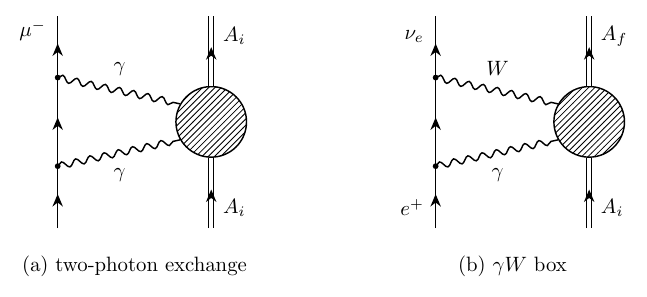}
    \caption{\label{fig:TPE_gammaW}
    Diagrammatic kinship of the two corrections reviewed here: (a) two-photon exchange in muonic atoms and (b) the $\gamma W$ box in superallowed $\beta$ decay, drawn for a $\beta^+$ transition such as $\carbonfermidecay$. Thin and double lines denote the leptons and the nucleus, respectively. The hatched blob is the same object in both diagrams, the generalized hadronic tensor, carrying two electromagnetic currents in (a) and one electromagnetic and one charged-weak current in (b).}
    \end{center}
\end{figure}

The theme of this review is that the leading nuclear-structure corrections entering the aforementioned flagship programs are two faces of the same theoretical coin. The relevant object is the forward electroweak current-current correlator with external nuclear states, i.e., the generalized hadronic tensor, as pictured in Figure~\ref{fig:TPE_gammaW}. In muonic atoms both currents are electromagnetic, and the correlator governs the TPE contribution to the Lamb shift. In superallowed $\beta$ decays, one current is electromagnetic and the other the weak-charged current, and the correlator constitutes the nuclear $\gamma W$ box correction. This kinship was recognized early in the modern reassessment of the radiative corrections, when the sum-rule and Lanczos techniques developed for muonic atoms were identified as the natural tools for the nuclear $\gamma W$ box~\cite{arXiv:1907.02164}. It extends to the entire computational chain: the spectral decomposition over nuclear intermediate states, the multipole expansion of the current operators, the Lanczos-based methods giving access to the required response functions and resolvents, and the order-by-order $\chi$EFT analysis underlying the uncertainty estimates. The dividends of a unified treatment are visible in two recent milestones, namely (i) the TPE corrections in muonic helium computed with Bayesian inference, which enabled the precise determinations of the helion and $\alpha$-particle charge radii and exonerated the nuclear-structure corrections in the helium isotope-shift puzzle~\cite{Lim25}, and (ii) one of the first \textit{ab-initio} computations of the nuclear $\gamma W$ box for the lightest superallowed transition, $\carbonfermidecay$~\cite{PRL.134.012501.2025}. The two programs are moreover coupled in practice and not only in formalism: the reference charge radii anchoring the superallowed analysis of the $^{26m}\mathrm{Al}\to{}^{26}\mathrm{Mg}$ transition are themselves extracted from muonic X-ray spectroscopy, so that the nuclear polarization correction propagates into the statistical rate function and, ultimately, into $V_{ud}$. A recent re-analysis along these lines reduced the CKM unitarity deficit by one standard deviation~\cite{PRR.7.L042002.2025}.

In this review, we concentrate on the nuclear-structure-dependent corrections entering the muonic-atom and superallowed programs with a particular focus on light nuclei, where wave-function-based \textit{ab-initio} methods provide numerically exact solutions of the many-body problem and rigorous uncertainty quantification is within reach. Heavier systems, for which this level of control is not yet available, are briefly discussed in connection with the recent experimental program in muonic X-ray spectroscopy~\cite{Oha24, Ung24, Bey25, Eiz26}. While not covered in this review, the same machinery underpins the analysis of $\beta$-spectra and angular correlations in searches for exotic currents~\cite{Gli22,Kin23}. We address the single-nucleon contributions to both observables only insofar as they enter the nuclear analysis~\cite{Car11,Ant22b}. Comprehensive accounts of muonic-atom spectroscopy and its theory may be found in Refs.~\cite{Ant22a,Pac23,Ji18}, of the superallowed data set in Ref.~\cite{PRC.102.045501.2020} and electroweak radiative corrections to $\beta$ decay in Ref.~\cite{ARNPS.74.2347.2024}, and of modern \textit{ab-initio} many-body methods in Ref.~\cite{Her20}.

The review is organized as follows. In Section~\ref{sec:common_theory}, we introduce the common theoretical framework: the generalized hadronic tensor, its spectral representation in terms of nuclear eigenstates, and the multipole decomposition of the electroweak current operators. In Section~\ref{sec:modern_ab_initio_methods}, we summarize the state-of-the-art in terms of \textit{ab-initio} theory: $\chi$EFT Hamiltonians and currents, the many-body and Lanczos-based methods used to compute the relevant amplitudes, and Bayesian uncertainty quantification. In Section~\ref{sec:muonic_atoms}, we review recent progress in muonic atoms, from the TPE formalism and the associated multipole sum rules to Bayesian-quantified predictions for muonic helium, their impact on the helium isotope-shift puzzle, and developments in heavy muonic atoms. In Section~\ref{section:superallowed}, we turn towards superallowed $\beta$ decays, specializing the hadronic tensor to the $\gamma W$ box, reviewing the first \textit{ab-initio} calculations, and discussing the implications for $V_{ud}$ and CKM unitarity. In Section~\ref{sec:conclusions_future_directions}, we conclude and discuss anticipated future directions.


\section{Common Theory of Nuclear-Structure Corrections}\label{sec:common_theory}

The leading nuclear-structure-dependent corrections introduced in Section~\ref{sec:introduction} -- the two-photon exchange in muonic atoms and the $\gamma W$ box in superallowed $\beta$ decays -- are both two-boson-exchange amplitudes in which the lepton lines and bosonic propagators factorize into process-dependent kinematic structures, while all nuclear structure is captured by a single object: the generalized hadronic tensor, a forward current-current correlator of electroweak operators evaluated with external nuclear states, nowadays accessible with \textit{ab-initio} many-body theory. Throughout, we work in natural units ($\hbar=c=1$) with unit-normalized nuclear states; matching to the relativistic normalization of amplitudes employed in $\beta$-decay literature is discussed in Section~\ref{section:superallowed}. In Section~\ref{sec:hadronic_tensor}, we define the generalized hadronic tensor and derive its spectral representation in terms nuclear eigenfunctions. In Section~\ref{sec:multipoles_current_operators}, we decompose the electroweak current operators into multipoles, the most pragmatic form in which nuclear amplitudes of currents are evaluated.


\subsection{The Hadronic Tensor and Current-Current Correlators}\label{sec:hadronic_tensor}

The probes considered in this review couple to the nucleus through the electromagnetic and the charged weak currents which, for the two lightest quark degrees of freedom, read
\begin{align}
    J^{\mu}_{ \mathrm{em} } = \frac{2}{3} \bar{u} \gamma^{\mu} u - \frac{1}{3} \bar{d} \gamma^{\mu} d
    \qquad \qquad
    J^{\dagger \, \mu}_{ \mathrm{W} } = \bar{d} \, \gamma^{\mu} \big( 1 - \gamma_5 \big) u
    \qquad .
\end{align}
At the energies relevant to nuclear structure, however, one does not work with quark degrees of freedom. The nuclear amplitudes of the above currents are evaluated with effective operators derived from chiral effective field theory ($\chi$EFT) and formulated in terms of nucleon and pion degrees of freedom; the operators are organized into one- and many-body contributions and are constructed consistently with the nuclear interaction~\cite{Kre20}, as discussed in Section~\ref{Section: 3.1}. The quark-level expressions fix the relevant charges, isospin structure, and vector or axial-vector nature of the probes that the effective operators realize at low energies. With the current operators understood in this effective sense, and introducing the momentum-space current $J_X^\mu(t,\mathbf q) = \int d^3x\, e^{-i\mathbf q\cdot\mathbf x}\,J_X^\mu(t,\mathbf x)$, we define the generalized non-relativistic hadronic tensor with unit-normalized external states as
\begin{equation}
    H_{XY}^{\mu\nu}(q;f,i)
    \equiv
    -i \int dt\, e^{iq_0 t}
    \bra{f}
    T\left\{J_X^\mu(t,\mathbf q)J_Y^\nu(0,-\mathbf q)\right\}
    \ket{i}
    +
    B_{XY}^{\mu\nu}(\mathbf q;f,i)
    \quad ,
    \label{eq:hadronic_tensor_definition_coordinate}
\end{equation}
where $X$ and $Y$ label the currents and external probes. For example, $X,Y=\mathrm{em}$ corresponds to the electromagnetic two-photon-exchange problem, while $X=\mathrm{em}$ and $Y=\mathrm{W}$ corresponds to the electroweak current-current correlator relevant to radiative corrections in $\beta$ decay. The term $B_{XY}^{\mu\nu}$ denotes the possible two-boson contact, or seagull, contributions. Finally, $\ket{i}$ and $\ket{f}$ are eigenstates of the nuclear Hamiltonian. Both current insertions in Eq.~(\ref{eq:hadronic_tensor_definition_coordinate}) are Fourier transformed with respect to their spatial arguments and carry opposite three-momenta, $\mathbf q$ and $-\mathbf q$, so that no net three-momentum is transferred to the external nucleus; the remaining integral is the transform with respect to the relative time separation of the two insertions. The overall normalization is a matter of convention and is chosen here such that the spectral representation provided below has the same structure as utilized in the muonic-atom two-photon-exchange formalism (in $\beta$-decay literature there is additional factor of $1/2$ in the normalization of the relativistic tensor).

For applications to nuclear many-body theory, it is useful to evaluate Eq.~(\ref{eq:hadronic_tensor_definition_coordinate}) in terms of nuclear eigenstates. We express the momentum-space current defined above in the Heisenberg picture with $ H_{\rm nuc} $ the nuclear Hamiltonian as
\begin{equation}
    J_X^\mu(t,\mathbf q)
    =
    e^{iH_{\rm nuc}t}J_X^\mu(\mathbf q)e^{-iH_{\rm nuc}t}
    \qquad \quad
    H_{\rm nuc}\ket{N}=E_N\ket{N}
    \quad .
    \label{eq:current_heisenberg_nuclear}
\end{equation}
One then expands the time-ordered product, resolves the intermediate nuclear spectrum, and performs the remaining $t$ integration to yield
\begin{eqnarray}
    H_{XY}^{\mu\nu}(q;f,i)
    &=&
    \sum_N
    \frac{
    \bra{f}J_X^\mu(\mathbf q)\ket{N}
    \bra{N}J_Y^\nu(-\mathbf q)\ket{i}
    }
    {E_f-E_N+q_0+i\epsilon}
    +
    \sum_N
    \frac{
    \bra{f}J_Y^\nu(-\mathbf q)\ket{N}
    \bra{N}J_X^\mu(\mathbf q)\ket{i}
    }
    {E_i-E_N-q_0+i\epsilon}
    \nonumber \\
    &&
    + \
    B_{XY}^{\mu\nu}(\mathbf q;f,i)
    \quad ,
    \label{eq:general_hadronic_tensor}
\end{eqnarray}
which is the form most amenable to nuclear response and sum rule calculations in \textit{ab-initio} theory. It also clarifies that all nuclear structure enters via the intermediate spectrum of $H_{\rm nuc}$ and the corresponding transition amplitudes of the electroweak current operators.

At this point, it is useful to clarify the meaning of ``forward''. In the language of scattering theory, a forward tensor refers to kinematics with no overall momentum transfer to the external target. This is manifest in Eq.~(\ref{eq:general_hadronic_tensor}), where the two current insertions carry opposite momenta, $\mathbf q$ and $-\mathbf q$ so that no net three-momentum is transferred to the external nucleus. For the electromagnetic two-photon-exchange problem in muonic atoms, the external initial and final nuclear states are identical, $\ket{i}=\ket{f}$, and the tensor is an elastic forward electromagnetic correlator. For superallowed $\beta$ decay, the external nuclear states are instead connected by the weak transition and are not identical. In that case, the corresponding object is a transition current-current correlator.

In the non-relativistic theory, the center-of-mass motion separates exactly from the intrinsic dynamics, and, since the two current insertions carry opposite three-momenta, momentum conservation fixes the center-of-mass momentum of the intermediate states to $\mp\mathbf{q}$ relative to the external nucleus. The eigenvalues $E_N$ entering the denominators of Eq.~(\ref{eq:general_hadronic_tensor}) then contain, in addition to the intrinsic energies of the nuclear eigenstate with mass $ M_{N} $, the center-of-mass energy $ \mathbf{q}^2 / 2 M_N $ of the recoiling intermediate nucleus. Further, all non-relativistic amplitudes of current operators reduce to intrinsic amplitudes. Whether this recoil shift of the pole positions is retained depends on the accuracy sought. It is kept explicitly in the $\gamma W$ box analysis of Section~\ref{section:superallowed}, whereas in the muonic-atom application it is suppressed by the lepton-to-nucleus mass ratio and does not enter the leading multipole sum rules of Section~\ref{sec:muonic_atoms}.

For the electromagnetic two-photon-exchange problem relevant to muonic atoms, one sets $X=Y=\mathrm{em}$ and $\ket{i}=\ket{f}=\ket{0}$. The generalized tensor then reduces to
\begin{eqnarray}
    H_{\mathrm{em}\,\mathrm{em}}^{\mu\nu}(q;0,0)
    &=&
    \sum_N
    \frac{
    \bra{0}J_{\mathrm{em}}^\mu(\mathbf q)\ket{N}
    \bra{N}J_{\mathrm{em}}^\nu(-\mathbf q)\ket{0}
    }
    {E_0-E_N+q_0+i\epsilon}
    +
    \sum_N
    \frac{
    \bra{0}J_{\mathrm{em}}^\nu(-\mathbf q)\ket{N}
    \bra{N}J_{\mathrm{em}}^\mu(\mathbf q)\ket{0}
    }
    {E_0-E_N-q_0+i\epsilon}
    \nonumber \\
    &&
    + \
    B_{\mathrm{em}\,\mathrm{em}}^{\mu\nu}(\mathbf q)
    \quad ,
    \label{eq:em_hadronic_tensor}
\end{eqnarray}
the elastic forward electromagnetic tensor evaluated with (the same) external nuclear ground states; this process is to be discussed in Section~\ref{sec:muonic_atoms}.

For the electroweak radiative corrections to superallowed $\beta$ decay, the relevant object contains one electromagnetic and one charged weak current. Taking $X=\mathrm{em}$ and $Y=\mathrm{W}$ in Eq.~(\ref{eq:general_hadronic_tensor}), one obtains
\begin{align}
    H_{\mathrm{em}\,\mathrm{W}}^{\mu\nu}(q;f,i)
    &=
    \sum_N
    \frac{
    \bra{f}J_{\mathrm{em}}^\mu(\mathbf q)\ket{N}
    \bra{N}J_{\mathrm{W}}^\nu(-\mathbf q)\ket{i}
    }
    {E_f-E_N+q_0+i\epsilon}
    +
    \sum_N
    \frac{
    \bra{f}J_{\mathrm{W}}^\nu(-\mathbf q)\ket{N}
    \bra{N}J_{\mathrm{em}}^\mu(\mathbf q)\ket{i}
    }
    {E_i-E_N-q_0+i\epsilon}
    \nonumber \\[0.15cm]
    & \qquad + \
    B_{\mathrm{em}\,\mathrm{W}}^{\mu\nu}(\mathbf q)
    \quad ,
    \label{section:hadronictensor:equation:emw_explicit}
\end{align}
where $\ket{i}$ and $\ket{f}$ denote the initial and final nuclear states connected by the weak transition. Note that these states are not necessarily the ground states of a given system. For spin-zero external states, as in superallowed $0^+\to0^+$ Fermi transitions, the vector-axial-vector part of this tensor can be projected onto a single scalar amplitude denoted by $T_3$. It is to be discussed in Section~\ref{section:superallowed} along with a comment on the nature of $ B_{\mathrm{em}\,\mathrm{W}}^{\mu \nu }(\mathbf{q}) $ in the context of the $\gamma W$ box correction.


\subsection{Multipoles of Current Operators}\label{sec:multipoles_current_operators}

The spectral representation of the hadronic tensor introduced in Section~\ref{sec:hadronic_tensor} is written in terms of amplitudes of momentum-space electroweak current operators. In practical \textit{ab-initio} calculations, these currents are usually expanded in multipoles and thus decomposed into a sum over rank-$J$ irreducible tensor operators. This provides a systematic expansion which separates away the angular dependence of the current and allows for a simple truncation scheme in the number of multipolar contributions. We separate the Minkowski-space current into its time and space components
\begin{equation}
    J_X^\mu(q)=\left(\rho_X(q),\mathbf J_X(q)\right)
    \quad ,
    \label{eq:current_charge_spatial_parts}
\end{equation}
where $\rho_X\equiv J_X^0$ is the nuclear charge-density operator and $\mathbf J_X$ is the nuclear current-density operator. The following decomposition applies equally to the electromagnetic and weak currents, albeit with proper accounting for $ \gamma_{5} $ and changes in the single-nucleon form factors. Following the standard multipole formalism for semi-leptonic weak and electromagnetic interactions with nuclei~\cite{Wal04, Don79}, and choosing the quantization axis along the $\mathbf{q}$ direction, the charge density and the spatial current decompose into irreducible tensor operators (thus with definite angular momentum and parity) as
\begin{equation}
    \rho_X(\mathbf q)
    =
    \sqrt{4\pi}\sum_{J=0}^{\infty}(-i)^J\sqrt{2J+1}\,M_{X,J0}(\mathbf{q})
    \quad ,
    \label{eq:charge_density_multipole_expansion}
\end{equation}
and
\begin{align}
    \mathbf J_X(\mathbf q)
    &=
    \sqrt{4\pi}\sum_{J=0}^{\infty}(-i)^J\sqrt{2J+1}\,
    L_{X,J0}(\mathbf{q})\,\hat{\mathbf e}_0(\hat{\mathbf q})
    \nonumber \\
    & \quad +
    \sqrt{2\pi}\sum_{\lambda=\pm1}\sum_{J=1}^{\infty}
    (-i)^J\sqrt{2J+1}
    \left[T^{\mathrm{el}}_{X,J\lambda}(\mathbf{q})-\lambda\,T^{\mathrm{mag}}_{X,J\lambda}(\mathbf{q})\right]
    \hat{\mathbf e}^{\,*}_{\lambda}(\hat{\mathbf q})
    \quad ,
    \label{eq:spatial_current_multipole_expansion}
\end{align}
where $\hat{\mathbf e}_0(\hat{\mathbf q})=\hat{\mathbf q}$ and $\hat{\mathbf e}_{\pm1}$ are the transverse spherical unit vectors. Eqs.~\eqref{eq:charge_density_multipole_expansion} and \eqref{eq:spatial_current_multipole_expansion} follow by insertion of the multipole expansion of the plane wave $e^{-i\mathbf q\cdot\mathbf x}$ in the Fourier transform of the current operators. The four operators $M_{X,JM}$, $L_{X,JM}$, $T^{\mathrm{el}}_{X,JM}$, and $T^{\mathrm{mag}}_{X,JM}$ are the Coulomb, longitudinal, transverse electric, and transverse magnetic multipoles, in the standard naming of Refs.~\cite{Wal04, Don79}.

The multipoles are projections of the nuclear charge and current densities onto a given basis of functions
\begin{gather}
    M_{X,JM}(\mathbf{q}) =\int d^3x \ j_J \big(\vert \mathbf{q} \vert \vert \mathbf{x} \vert \big) \ Y_{JM}(\hat{\mathbf x})\,\rho_X(\mathbf x) \quad ,\\
    L_{X,JM}(\mathbf{q}) =\frac{i}{ \vert \mathbf{q} \vert } \int d^3x \ \Big[\nabla\, j_J \big(\vert \mathbf{q} \vert \vert \mathbf{x} \vert \big) \ Y_{JM}(\hat{\mathbf x})\Big]\cdot\mathbf J_X(\mathbf x) \quad ,\\
    T^{\mathrm{el}}_{X,JM}(\mathbf{q}) =\frac{1}{ \vert \mathbf{q} \vert } \int d^3x \ \Big[\nabla\times j_J \big(\vert \mathbf{q} \vert \vert \mathbf{x} \vert \big) \ \mathbf Y_{JJM}(\hat{\mathbf x}) \Big] \cdot\mathbf J_X(\mathbf x) \quad ,\\
    T^{\mathrm{mag}}_{X,JM}(\mathbf{q}) =\int d^3x \ j_J \big(\vert \mathbf{q} \vert \vert \mathbf{x} \vert\big) \ \mathbf Y_{JJM}(\hat{\mathbf x})\cdot\mathbf J_X(\mathbf x) \quad ,
    \label{eq:multipole_operator_definitions}
\end{gather}
where $j_J$ is a spherical Bessel function, and $Y_{JM}$ and $\mathbf Y_{JJM}$ are the spherical and vector spherical harmonics, respectively. The single-particle reduced amplitudes of the multipole operators are known and tabulated in the harmonic-oscillator basis~\cite{Don79} and available symbolically for arbitrary quantum numbers~\cite{Hax08}. The four operators above exhaust the multipole content of the electromagnetic current. However, since the charged-weak current has a $ V - A $ structure, there exists an additional axial-vector multipole for each vector-current multipole listed above. We denote these multipoles by the superscript ``$5$'', for example, $T^{5,\mathrm{el}}_{\mathrm{W},JM}$ or $T^{5,\mathrm{mag}}_{\mathrm{W},JM}$, and note that they carry parity opposite to their vector counterparts and enter the $\gamma W$ box multipole sum of Section~\ref{section:superallowed}.

For conserved vector currents, the Coulomb and longitudinal multipoles are related by the continuity equation. In momentum space, current conservation implies
\begin{equation}
    q_0 \ \rho_X(\mathbf q)=\mathbf q\cdot\mathbf J_X(\mathbf q)
    \quad ,
    \label{eq:continuity_equation_momentum_space}
\end{equation}
where $q_0$ is fixed by the amplitude under consideration. This fixes the longitudinal multipole in terms of the Coulomb one as
\begin{equation}
    L_{X,JM}(q)=\frac{q_0}{q}\,M_{X,JM}(q)
    \quad .
    \label{eq:coulomb_longitudinal_relation}
\end{equation}
This relation is routinely used to simplify electromagnetic response calculations. The axial-vector current, by contrast, is not conserved and the corresponding longitudinal multipole must be retained explicitly~\cite{Wal04}.


\section{Modern \textit{Ab-Initio} Methods}\label{sec:modern_ab_initio_methods}

The common formalism introduced in Section~\ref{sec:common_theory} reduces the hadronic-structure-dependent corrections of interest to amplitudes and response functions involving electroweak currents in nuclei. This section summarizes the ingredients from \textit{ab-initio} theory needed to evaluate these quantities in practical situations. In Section~\ref{Section: 3.1}, we discuss nuclear Hamiltonians, electroweak current operators, and the many-body methods used to compute the relevant amplitudes. Section~\ref{Section: 3.2} then describes how Lanczos techniques can be used to access the response functions and sum rules entering the hadronic tensor. Finally, Section~\ref{Section: 3.3} reviews recent developments in uncertainty quantification, with particular emphasis on the propagation of $\chi$EFT truncation errors.


\subsection{Nuclear Hamiltonians, Electroweak Operators and \textit{Ab-initio} Theory}\label{Section: 3.1}

Microscopic inputs to the hadronic tensor consist of a chosen nuclear Hamiltonian, a basis of electroweak current operators, and a many-body method capable of evaluating the required ground-state and transition amplitudes. In an \textit{ab-initio} approach, these ingredients are constructed within a common theoretical framework and are then applied to observables for a given process without further adjustment. This consistency is essential for the applications discussed in this review.

Historically, quantitative nuclear-structure calculations relied heavily on high-precision phenomenological interactions. The development of accurate nucleon-nucleon (NN) potentials, such as Argonne v18 (AV18)~\cite{Wir95} and CD-Bonn~\cite{Mac01}, supplemented by phenomenological three-nucleon (3N) interactions such as the Urbana IX and Illinois models~\cite{Pud95,Pie01}, enabled a highly successful description of the properties of light nuclei. Constructed to reproduce NN scattering data and selected few-body observables with high accuracy, they established the predictive power of modern few-body calculations. Their empirical nature, however, left them without a systematic low-energy expansion of Quantum Chromodynamics (QCD), a controlled hierarchy of many-body forces, or a natural prescription for propagating theoretical uncertainties.

A major conceptual advance came with Weinberg's proposal to formulate nuclear forces within effective field theory~\cite{Wei90,Wei91,Wei92}. Chiral effective field theory
($\chi$EFT) has since become the standard framework for constructing nuclear Hamiltonians consistent with the symmetries of low-energy QCD~\cite{Epe09,Mac11,Ham20}. In this
approach, the relevant degrees of freedom are nucleons and pions, while unresolved short-distance physics is encoded in low-energy constants (LECs) that are constrained by experimental data. The resulting nuclear Hamiltonian takes the schematic form
\begin{equation}
    H_{\rm nuc}
    =
    T_{\rm rel}
    +
    V_{\rm NN}
    +
    V_{\rm 3N}
    +
    V_{\rm 4N}
    +\cdots
    \quad ,
    \label{eq:nuclear_hamiltonian_general}
\end{equation}
where $T_{\rm rel}$ is the relative kinetic energy and the nuclear interactions are organized according to the chiral expansion. In Weinberg's power counting the leading order Hamiltonian only contains NN forces, 3N forces emerge at next-to-next-to-leading order (N$^2$LO), and 4N forces arise at higher orders~\cite{Epe09, Mac11}. Modern chiral interactions have reached a high level of sophistication. Two-nucleon forces have been derived to high orders in the chiral expansion, including N$^4$LO and partial N$^5$LO contributions~\cite{Epe15,Epe20}. At the same time, the role of the $\Delta(1232)$ nucleon resonance has motivated the development of $\Delta$-full formulations, in which the $\Delta$ is included as an explicit degree of freedom. This can improve the convergence pattern of the chiral expansion by incorporating important intermediate-range physics at lower orders~\cite{Epe09,Mac11,Pia15}.

Despite this progress, the formulation and use of chiral interactions involve important conceptual and practical issues. Chiral potentials are singular at short distances and must be regularized before they can be iterated in the Schr\"odinger equation. In the standard implementation of Weinberg's power counting, the cutoff is kept finite and chosen to be of the order of the EFT breakdown scale. Observables therefore retain a residual cutoff dependence, which should decrease systematically as higher orders are included. An alternative strategy emphasizes strict renormalization-group invariance and treats sub-leading interactions perturbatively after solving the leading-order problem non-perturbatively~\cite{Ham20,Thi26,Thi26_phdthesis}. The relation between these approaches, and the extent to which they provide controlled order-by-order convergence in nuclei, remains an active area of research.

The regulator structure of chiral interactions also matters for many-body calculations. Many chiral potentials are naturally formulated in momentum space and are non-local. This can be challenging for coordinate-space methods, particularly Quantum Monte Carlo (QMC) approaches. A significant effort has therefore been devoted to constructing local or maximally local chiral interactions suitable for coordinate-space calculations~\cite{Gez14,Lyn16,Pia20,Som24}. By exploiting regulator choices and Fierz rearrangement freedom, local NN interactions have been developed to high chiral orders, and local 3N forces have been implemented through N$^2$LO~\cite{Lyn16,Som24}. These interactions have enabled QMC calculations of light nuclei and neutron matter, and they are also useful in other coordinate-space approaches, such as hyperspherical harmonics calculations \cite{Lim21}.

For the observables discussed in this review, the Hamiltonian alone is not sufficient. The hadronic tensor in Eq.~(\ref{eq:general_hadronic_tensor}) also depends on electromagnetic and weak current operators. These operators must be constructed consistently with the nuclear interaction. In $\chi$EFT, electroweak currents are organized in the same low-energy expansion as the nuclear forces and contain both
one-body and many-body contributions~\cite{Kre20}. Schematically,
\begin{equation}
    J_X^\mu
    =
    \sum_i J_{X,i}^{\mu,{\rm 1b}}
    +
    \sum_{i<j} J_{X,ij}^{\mu,{\rm 2b}}
    +
    \cdots
    \quad ,
    \label{eq:current_operator_expansion}
    \end{equation}
where $X=\mathrm{em}$ (electromagnetic) or $\mathrm{W}$ (weak), and the ellipsis denotes higher-body current operators. The one-body term, in which the external probe couples to individual nucleons, contains the leading impulse approximation currents. Two-body currents encode mechanisms in which the probe couples to correlated nucleon pairs, and they include meson-exchange effects. These terms are required by current conservation for consistency with the nuclear Hamiltonian.

The quantitative impact of two-body electroweak currents has been demonstrated in a variety of few-body observables. For many weak transitions in light nuclei, the one-body impulse approximation accounts for the dominant contribution, while two-body axial currents provide smaller but non-negligible corrections~\cite{Pas18,Kin20,Kin23}. Some transitions exhibit enhanced sensitivity: in selected $A=8$ $\beta$ decays~\cite{Kin20}, the leading one-body Gamow-Teller matrix element is suppressed by approximate Wigner spin-isospin symmetry, making the relative effect of two-body currents unusually large~\cite{Lim26}.

Once the Hamiltonian and electroweak current operators are specified, the remaining task is to solve the nuclear many-body problem and evaluate the amplitudes entering the spectral representation of the hadronic tensor. In light nuclei, methods such as Quantum Monte Carlo~\cite{Car15,Lyn17,Lon18}, Hyperspherical Harmonics (HH)~\cite{Viv06,Mar20}, their variants such as the Effective Interaction Hyperspherical Harmonics (EIHH)~\cite{Bar00,Lim21}, and the No-Core Shell Model (NCSM)~\cite{Bar13} can provide highly accurate solutions of the Schr\"odinger equation and detailed access to nuclear correlations. In medium-mass nuclei, many-body frameworks such as Coupled Cluster (CC)~\cite{Hag14} and In-Medium Similarity Renormalization Group (IMSRG)~\cite{Her16} extend the reach of \textit{ab-initio} calculations to heavier systems via controlled, polynomial-scaling approximations. A broad overview of modern \textit{ab-initio} methods may be found in Ref.~\cite{Her20}.


\subsection{The Lanczos Algorithm}\label{Section: 3.2}

In this section we review the Lanczos methods fundamental to calculations in configuration interaction theory. We will denote a general discretized Hamiltonian represented in a truncated configuration space basis of dimension $ d $ as $ \bar{H} \equiv \bar{H}(d) $ and note that $ \lim_{ d \rightarrow \infty } \bar{H}(d) = H $ in the weak sense (meaning the spectrum $ \sigma(H) $ is recovered). Then,
\begin{align}
    \bar{H} ( d ) = \begin{pmatrix}
        \ \bra{ \psi_{1}^{A} } \, H \, \ket{ \psi_{1}^{A} }
        & \cdots &
        \bra{ \psi_{1}^{A} } \, H \, \ket{ \psi_{d}^{A} } \ \\
        \ \vdots & \ddots & \vdots \ \\
        \ \bra{ \psi_{d}^{A} } \, H \, \ket{ \psi_{1}^{A} } & \cdots & \bra{ \psi_{d}^{A} } \, H \, \ket{ \psi_{d}^{A} } \
    \end{pmatrix}
    \quad ,
\end{align}
and the work to be done is that of matrix algebra; given the typical dimensions $ d $, one must apply approximate diagonalization schemes to construct the eigensolutions of $ \bar{H}(d) $.

For a given Hermitian operator, the most powerful such approach is the Lanczos algorithm~\cite{JRNBSB.45.255282.1950, SIAM.Komzsik.2003} which is premised on the existence of a special basis in which the matrix $ \bar{H}(d) $ has tri-diagonal form. A Krylov method, the Lanczos algorithm iteratively generates an orthonormal basis from recursive application of the desired operator on a random starting vector, called the \textit{pivot}, which has non-zero overlap with the desired part of the spectrum. In practice, the number of algorithm iterations $ N_{ \mathrm{iter} } $ is guided by the characteristics of the spectrum in question, e.g.,
(i) the eigenvalue density,
(ii) the number of desired converged eigensolutions,
(iii) the number of degrees of freedom, and
(iv) the dimension $ d $.
While completion of the algorithm with $ N_{ \mathrm{iter} } = d $ is typically not possible, the remarkable advantage of the Lanczos algorithm is that one needs a number of iterations far less than the dimension of the matrix in most scenarios of interest, i.e., $ N_{ \mathrm{iter} } \ll d $. The conditions on convergence were already well-understood in his time by Lanczos~\cite{JRNBSB.45.255282.1950}, yet more recent studies have elucidated further details on bounds of the Lanczos-type methods, see Ref.~\cite{LAA.34.235258.1980} and references therein as well as Ref.~\cite{beresfordn.1995}.

Formally initialized from a random, normalized pivot vector $ \ket{ \eta_0 } $ -- though vectors bootstrapped from lower-dimensional configuration spaces offer a pragmatic convergence-accelerating alternative -- we generate a sequence of Lanczos basis vectors $ \{ \ket{ \eta_{i} }\} $ for a Hermitian operator $ \bar{H} $ as
\begin{align*}
    \bar{H} \, \ket{ \eta_{ 0 } } &= \alpha_0 \, \ket{ \eta_{ 0 } } + \beta_{ 0 } \, \ket{ \eta_{ 1 } } \\
    \bar{H} \, \ket{ \eta_{ 1 } } &= \beta_{ 0 } \, \ket{ \eta_{ 0 } } + \alpha_1 \, \ket{ \eta_{ 1 } } + \beta_{ 1 } \, \ket{ \eta_{ 2 } } \\
    \bar{H} \, \ket{ \eta_{ 2 } } &= \qquad \qquad \, \beta_{ 1 } \, \ket{ \eta_{ 1 } } + \alpha_2 \, \ket{ \eta_{ 2 } } + \beta_{ 2 } \, \ket{ \eta_{ 3 } } \\
    \bar{H} \, \ket{ \eta_{ 3 } } &= \qquad \qquad \qquad \qquad \, \, \beta_{ 2 } \, \ket{ \eta_{ 2 } } + \alpha_3 \, \ket{ \eta_{ 3 } } + \beta_{ 3 } \, \ket{ \eta_{ 4 } } \\
    & \ \,\, \vdots
\end{align*}
where each new vector in the sequence is constructed as explicitly orthonormal to its two leftmost neighbors (further re-orthogonalization against the whole sequence via the Gram-Schmidt procedure is often necessary) as
\begin{gather}
    \beta_{i} \, \ket{ \eta_{ i+1 } } = \bar{H} \, \ket{ \eta_{i} } - \alpha_{i} \, \ket{ \eta_{i} }
    - \beta_{i-1} \, \ket{ \eta_{ i-1 } }
    \quad , \\[0.5cm]
    \alpha_{i} = \bra{ \eta_{i} } \, \bar{H} \, \ket{ \eta_{i} }
    \qquad \qquad
    \beta_{i}^{2} = \Big\vert \Big\vert \ \big( \bar{H} - \alpha_{i} \mathds{1} \big) \ket{ \eta_{i} }
    - \beta_{i-1} \, \ket{ \eta_{ i-1 } } \ \Big\vert \Big\vert_{ L^{2} }^{2}
    \quad .
\end{gather}
The Lanczos basis spans the Krylov subspace generated by the pivot vector and $ \bar{H} $, that is,
\begin{align}
    \mathcal{K}_{ N_{ \mathrm{iter} } } \big( \eta_0, \bar{H} \big)
    &\equiv \mathrm{span} \Big\{ \, \ket{ \eta_0 }, \, \bar{H} \, \ket{ \eta_0 }, \, \bar{H}^{2} \, \ket{ \eta_0 }, \, \dots, \, \bar{H}^{ N_{ \mathrm{iter} } - 1 } \, \ket{ \eta_0 } \, \Big\}
    \quad , \nonumber
\end{align}
and casts an $ N_{ \mathrm{iter} } \times N_{ \mathrm{iter} } $ sub-block of $ \bar{H} $ into tri-diagonal form, which we denote as $ \bar{H}_{ \mathrm{L} } $. Thus the Rayleigh-Ritz eigensolutions of $ \bar{H} $ are given entirely in terms of the generated set of coefficients $ \{ \alpha_i, \, \beta_i \} $ and Lanczos vectors $ \{ \ket{ \eta_{i} }\} $ as
\begin{gather}\label{section:lanczos:equation:diagonalize}
    E \doteq P^{ -1 } \bar{H}_{ \mathrm{L} } \, P
    \qquad \qquad
    \ket{ \Psi_{i} } \doteq \sum_{ j = 1 }^{ N_{ \mathrm{iter} } } \, \overlap{ \eta_{j} }{ \Psi_{i} } \; \ket{ \eta_{j} }
    = \sum_{ j = 1 }^{ N_{ \mathrm{iter} } } \, \big( P^{-1} \big)_{ij} \, \ket{ \eta_{j} }
    = \sum_{ j = 1 }^{ N_{ \mathrm{iter} } } \, P_{ji} \, \ket{ \eta_{j} }
    \quad ,
\end{gather}
where $ P $ denotes the $ N_{ \mathrm{iter} } \times N_{ \mathrm{iter} } $ unitary matrix containing the coordinates of the Rayleigh-Ritz eigenvectors in the Lanczos basis.


Let us now consider a problem in which we have a bounded Hamiltonian operator $ \bar{H} $ and we would like to evaluate amplitudes of the following form
\begin{gather}
    \mathcal{A}_{ fi } = \bra{ \Psi_{f} } \, \bar{O}_{ 2 } \, ( z - \bar{H} )^{ -1 } \, \bar{O}_{ 1 } \, \ket{ \Psi_{i} }
    = \bra{ \Psi_{f} } \, \bar{O}_{ 2 } \, \ket{ \Psi_{ \mathrm{R} } }
    \quad , \label{section:lanczos:equation:amplitude} \\[0.33cm]
    ( z - \bar{H} ) \, \ket{ \Psi_{ \mathrm{R} } } = \bar{O}_{1} \, \ket{ \Psi_{i} }
    \quad , \label{section:lanczos:equation:ISE}
\end{gather}
where $ z \in \mathds{C} \setminus \sigma (\bar{H}) $. Without loss of generality, we will assume that $ \bar{O}_{1} $ and $ \bar{O}_{2} $ are irreducible spherical tensor operators which carry good spin-parity numbers, and we note that any relevant operator may be projected onto good spin-parity via an expansion akin to the multipole decomposition given in Section~\ref{sec:multipoles_current_operators}. One clearly recognizes that Eq.~\eqref{section:lanczos:equation:ISE} is an inhomogeneous Schr{\"o}dinger equation with a source given by the action of $ \bar{O}_{1} $ on the initial state and, almost exactly as in the homogeneous case, the solutions may be constructed via the Lanczos algorithm. Remarkably though, the admitted solutions will contain information about transitions to the entire spectrum relevant to construction of the Green function in the desired amplitude
\begin{align}\label{section:lanczos:equation:gfamplitude}
    \mathcal{A}_{ fi }
    = \sumintegral_{n} \ \frac{ \bra{ \Psi_{f} } \, \bar{O}_{ 2 } \, \ket{ \Psi_{n} } \bra{ \Psi_{n} } \, \bar{O}_{ 1 } \ket{ \Psi_{i} } }
    { z - \omega_{n} }
    \quad ,
\end{align}
where $ \omega_{n} $ denotes the eigenvalue of $ \bar{H} $ associated with the eigensolution $ \ket{ \Psi_{n} } $. The formal aspects of this problem have been studied over several decades, e.g., see the early ideas developed for the ``recursion method'' as applied in condensed matter physics in the works of Haydock, Heine and Kelly~\cite{JPA.7.17.1974, SSP.35.215294.1980, SSP.35.1127.1980, SSP.35.295383.1980, JPCSSP.5.2845.1972}, as well as the $ 1987 $ proceedings in Ref.~\cite{SSBM.58.2012}. Moreover, the \textit{ab-initio} nuclear theory community has exploited this technique over the last two decades~\cite{FBS.33.259276.2003, PRC.72.065501.2005, PRC.89.064317.2014, PRA.102.052828.2020, PRC.104.025502.2021}, with a recent application to electroweak radiative corrections in superallowed beta decays~\cite{PRL.134.012501.2025}.

Modification to the Lanczos algorithm occurs due to the presence of the inhomogeneity; rather than a random pivot, one sets the starting vector to be the normalized constraint
\begin{align}
    \ket{ \eta_{0} } = \frac{1}{ \sqrt{ \bra{ \Psi_{i} } \, \bar{O}_{1}^{\dagger} \, \bar{O}_{1} \, \ket{ \Psi_{i} } } } \
    \bar{O}_{1} \, \ket{ \Psi_{i} }
    \quad .
\end{align}
Then, following Ref.~\cite{RMP.66.763840.1994}, we may explicitly construct the desired transition amplitudes in Eq.~\eqref{section:lanczos:equation:gfamplitude} by expansion of the intermediate eigensolutions $ \Psi_{n} $ in the Lanczos basis, by definition given by $ \ket{ \Psi_{n} } = \sum_{m} \, c_{nm} \, \ket{ \eta_{m} } $. The relevant amplitudes are then
\begin{gather}
    \bra{ \Psi_{n} } \, \bar{O}_{1} \, \ket{ \Psi_{i} }
    = \sqrt{ \bra{ \Psi_{i} } \, \bar{O}_{1}^{ \dagger } \, \bar{O}_{ 1 } \, \ket{ \Psi_{i} } } \
    \bra{ \eta_{n} } \, P^{\dagger} \, \ket{ \eta_{0} }
    \quad , \nonumber \\[0.3cm]
    \bra{ \Psi_{f} } \, \bar{O}_{2} \, \ket{ \Psi_{n} }
    = \sqrt{ \bra{ \Psi_{f} } \, \bar{O}_{2}^{ \dagger } \, \bar{O}_{ 2 } \, \ket{ \Psi_{f} } } \
    \sum_{m} \, \bra{ \eta_{m} } \, P \, \ket{ \eta_{n} } \ \overlap{ \zeta_{0} }{ \eta_{m} }
    \quad ,
\end{gather}
where we recall $ P $ from Eq.~\eqref{section:lanczos:equation:diagonalize} to be the unitary matrix which diagonalizes the $ N_{\mathrm{iter}} \times N_{\mathrm{iter}} $ tri-diagonal sub-block of the Hamiltonian represented in the Lanczos basis.

In fact, one may trivially evaluate generalized quantities with the above amplitudes of the form
\begin{align}
    I_{fi} &= \int dz \ \bra{ \Psi_{f} } \, \bar{O}_{2} \ f ( z, \bar{H} ) \ \bar{O}_{1} \, \ket{ \Psi_{i} }
    = \sumintegral_{n} dz \ f ( z, \omega_{n} ) \
    \bra{ \Psi_{f} } \, \bar{O}_{2} \, \ket{ \Psi_{n} } \ \bra{ \Psi_{n} } \, \bar{O}_{1} \, \ket{ \Psi_{i} }
    \quad ,
\end{align}
subject to the analytic properties of the function $ f $ and the choice of $ N_{\mathrm{iter}} $~\cite{JPCSSP.5.2845.1972, RMP.66.763840.1994, JMP.16.462474.1975}; by construction, the first $ N_{\mathrm{iter}} - 1 $ terms of the Taylor series of $ f(z,\bar{H}) $ are exact. Note that the amplitudes $ \mathcal{A}_{fi} $ used to motivate this section are merely a simple case with the choice of analytic function $ f(z, \bar{H}) = (z - \bar{H})^{-1} $ on the domain $ \mathds{C} \setminus \sigma (\bar{H}) $.


\subsection{Uncertainty Quantification}\label{sec:uncertainty_quantification}\label{Section: 3.3}

A distinguishing advantage of chiral effective field theory ($\chi$EFT) is that it provides not only a systematic framework for constructing nuclear forces and electroweak operators, but also a statistically rigorous setting for estimating theoretical uncertainties. In practical applications, the dominant source of theory uncertainty is often the truncation of the chiral expansion at finite order. Such errors were historically estimated heuristically, for example from the size of the last computed correction or from variations of the regulator scale. Use of Bayesian statistics has since emerged as the gold standard for quantifying EFT truncation errors in a way that is transparent and statistically interpretable~\cite{Fur15,Mel19,Lim23,Eks20}.

The starting point is the EFT expansion of a generic observable $X$, computed at chiral order $k$. The prediction $X^{(k)}$ can be parameterized in terms of a dimensionless expansion parameter $Q$, typically taken as $ Q = \max(p,m_\pi)/\Lambda_b $, where $p$ denotes a characteristic momentum scale of the process and $\Lambda_b$ is the EFT breakdown scale~\cite{Fur15,Mel17,Mel19,Sch09,Lim23}. One then writes
\begin{equation}
    X^{(k)}
    =
    X_{\mathrm{ref}}
    \sum_{n=0}^{k} c_n Q^n
    \qquad \quad
    X = X^{(k)} + \Delta_k
    \quad ,
    \label{eq:uq_eft_expansion}
\end{equation}
where $X_{\mathrm{ref}}$ sets the overall dimensionful scale, the $c_n$ are dimensionless expansion coefficients, and
\begin{equation}
    \Delta_k
    =
    X_{\mathrm{ref}}
    \sum_{n=k+1}^{\infty} c_n Q^n
    \quad ,
\end{equation}
denotes the residual truncation error. The coefficients $c_n$ can be extracted recursively from the order-by-order results. The core EFT assumption is that, once the relevant scales have been factored out, these coefficients are natural, i.e., typically of order unity.

Bayesian uncertainty quantification translates this naturalness assumption into prior probability distributions for the unknown coefficients, often formulated in terms of an underlying scale parameter $\bar c$ that characterizes their typical size. Bayes' theorem is then used to update this prior information using the known coefficients $\{ c_0, \, \dots, \, c_k \}$. Marginalizing over $\bar c$ yields a posterior probability distribution for the truncation error $\Delta_k$. A major result of this framework was that several older, truncation-error prescriptions were able to be understood as particular, and sometimes overly restrictive, choices of a Bayesian prior~\cite{Fur15}. The resulting interval acquires a strict degree-of-belief (DOB) interpretation: conditional on the EFT model and the adopted priors, a $68\%$ DOB interval contains the exact result with $68\%$ probability.

For observables that depend on external kinematic variables, e.g., scattering energy, scattering angle, or momentum transfer, uncertainties at neighboring points are generally correlated. A point-wise application of the Bayesian framework ignores this information and may therefore produce unphysical, jagged uncertainty bands. This issue was addressed in Ref.~\cite{Mel19}, where the EFT coefficients $c_n(x)$ were modeled as draws from a Gaussian process. The resulting formalism incorporates a covariance kernel that encodes the expected smoothness and correlation structure of the observable across the relevant domain. This yields statistically correlated uncertainty bands and a full covariance matrix for use in fits, hypothesis testing, and diagnostics of the EFT convergence pattern.

While the quantification of truncation errors has seen widespread adoption, a complete theoretical uncertainty quantification (UQ) framework must also account for the statistical uncertainties in the determination of the low-energy constants (LECs) of $\chi$EFT. The LECs are typically calibrated to experimental data, such as nucleon-nucleon scattering phase shifts and selected few-nucleon observables. Bayesian parameter estimation provides a rigorous methodology for extracting the joint posterior distribution of these parameters, including their uncertainties and correlations~\cite{Wes19,Sve23,Wes21,Eks20}. Propagating these parameter uncertainties to observables requires repeated solution of the many-body Schr\"odinger or Lippmann-Schwinger equation and is therefore computationally demanding. For this reason, modern nuclear theory has increasingly turned to reduced-order models, or emulators, as part of a broader statistical-computing strategy for uncertainty quantification \cite{Eks20}. In this context, eigenvector continuation (EC), originally introduced as a variational method for tracking eigenvectors in parameter-dependent Hamiltonians \cite{Fra18}, has emerged as a particularly promising emulator strategy for nuclear many-body observables, providing efficient and accurate emulation in multi-nucleon systems~\cite{Kon20,Dja22,Wes21}. Rigorous Bayesian sampling of the LECs has not yet been applied to nuclear polarizabilities in muonic atoms or to radiative corrections in superallowed $\beta$ decays; coupling EC-based emulators to electroweak transition amplitudes is the natural next step.

For the applications discussed in this review, the principal source of EFT uncertainty stems from the truncation of the chiral expansion in the nuclear Hamiltonian and in the electroweak current operators, which have to be propagated through the many-body calculation to the final response functions and transition amplitudes~\cite{Lim23,Ach23}.


\section{Recent Progress in Muonic Atoms}
\label{sec:muonic_atoms}

A muonic atom is a hydrogen-like bound system in which a negative muon replaces the atomic electron. Because the muon is about $207$ times heavier than the electron, its Bohr radius is correspondingly smaller and its wavefunction overlaps far more extensively with the nuclear charge distribution. Nuclear-structure effects in muonic atoms scale as $(m_\mu/m_e)^3$ relative to ordinary atoms and are therefore enormously enhanced, turning muonic atoms into precision probes of nuclear structure~\cite{Ant22a}. This sensitivity is the basis of the experimental program of the CREMA collaboration at the Paul Scherrer Institute, which has measured the $2S$--$2P_{1/2}$ Lamb shift by laser spectroscopy in muonic hydrogen~\cite{Poh10,Ant13}, muonic deuterium~\cite{Poh19}, and the muonic helium ions $\mu^3\mathrm{He}^+$ and $\mu^4\mathrm{He}^+$~\cite{Kra21,Sch25}.

\subsection{Physical context and observables}
\label{sec:muonic_physical_context}

The extraction of nuclear radii from spectroscopic measurements in atomic systems originates from a careful comparison with precise theoretical predictions. The measured Lamb shift in a muonic atom can be parametrized as
\begin{equation}
\delta_{\text{LS}} = \delta_{\text{QED}} + \mathcal{A}_{\text{OPE}}\, r_c^2 + \delta_{\text{NS},\mu} \ ,
\label{eq:muonic_lamb_shift}
\end{equation}
where $\delta_{\text{QED}}$ collects the purely quantum-electrodynamical and recoil contributions, see Ref.~\cite{Pac23} for a recent summary of these contributions. The second term is the leading finite-size effect: it arises from one-photon exchange with the nuclear charge form factor inserted at the photon--nucleus vertex, and is proportional to the mean-square charge radius $r_c^2$ through the known coefficient $\mathcal{A}_{\text{OPE}}\simeq m_r^3(Z\alpha)^4/12$, with $m_r$ the muon--nucleus reduced mass. The charge radius is the quantity of interest and is extracted by equating Eq.~(\ref{eq:muonic_lamb_shift}) to the spectroscopic measurement once $\delta_{\text{QED}}$ and $\delta_{\text{NS},\mu}$ are known.

The last term, $\delta_{\text{NS},\mu}$, includes nuclear structure effects which require knowledge of the full nuclear dynamics; the subscript $\mu$ distinguishes it from the analogous nuclear-structure correction $\delta_{\text{NS}}$ entering superallowed $\beta$ decay (Section~\ref{section:superallowed}). Its leading-order contribution is the radiative two-photon-exchange nuclear-structure correction $\delta_{\text{TPE}}$, of order $(Z\alpha)^5$, see Figure~\ref{fig:TPE}. While $\delta_{\text{QED}}$ is known to high accuracy and the spectroscopic measurements are extremely precise, $\delta_{\text{TPE}}$ is more difficult to calculate; it is by far the dominant source of uncertainty in the extracted radii~\cite{Ji18,Pac23,Ant22a}. 

Over the years, several groups have focused on the evaluation of these nuclear polarizability effects. In muonic hydrogen, one generally follows either a chiral-perturbation-theory approach, in its heavy-baryon~\cite{Nev08,Pes14,Bir12} or manifestly covariant~\cite{Ala14} formulation, or a data-driven approach, in which the two-photon-exchange correction is evaluated through dispersion relations using measured proton form factors and structure functions~\cite{Car11,Ant22b}. While the data-driven method requires subtracting an unknown function~\cite{Ant22b}, effective field theories (EFTs) avoid this subtraction ambiguity altogether. Instead, EFTs introduce low-energy constants (LECs) that must be determined through matching with experimental data or the underlying fundamental theory before making predictive calculations. In muonic deuterium and heavier systems different techniques must be applied, since one deals with very different energy scales and the most relevant physics is captured by the inter-nucleon interaction. Owing to the absence of three-nucleon forces and the simplicity of the system, works on muonic deuterium -- based on pioneering analytical evaluations and potential models~\cite{Fri13,Pac11,Pac15a}, chiral and pionless effective field theories~\cite{Her14,Her18,Her19,Kal19,Ach21,Len22,Len22b}, and data-driven dispersive analyses~\cite{Car14} -- achieved remarkable precision in the determination of the nuclear structure effects, whereas in muonic tritium~\cite{Nev16,Ji18} and in muonic helium ions~\cite{Ji13,Nev16,Ji18,Lim22} the precision of the calculated $\delta_{\text{NS},\mu}$ is at the level of a few percent.

Muonic helium has proved a particularly rich testing ground from a nuclear-structure perspective, for two reasons. The measurement of the muonic Lamb-shift in $\mu^4\mathrm{He}^+$ and $\mu^3\mathrm{He}^+$ gave the most precise extraction of the corresponding nuclear radii to date~\cite{Kra21,Sch25}, while the comparison of the helion ($^3\mathrm{He}$) and $\alpha$-particle ($^4\mathrm{He}$) radii confronts spectroscopy in muonic and ordinary atoms and gave rise to the helium isotope-shift puzzle -- recently resolved -- discussed in Section~\ref{sec:muonic_helium_isotope_shift_puzzle}. Reducing the theoretical uncertainty on $\delta_{\text{TPE}}$ with controlled \textit{ab-initio} methods is therefore the key to exploiting the experimental precision, and is the central theme of the remainder of this section.

\subsection{Two-photon-exchange amplitude in muonic atoms}
\label{sec:muonic_tpe_amplitude}

\begin{figure}[t]
\begin{center}
\includegraphics[width=0.9\linewidth]{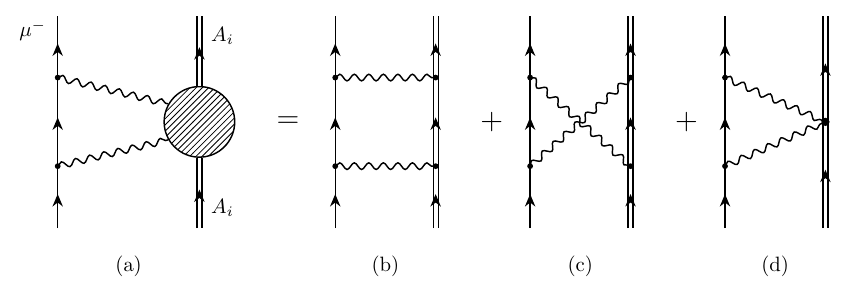}
\caption{(a) Self-energy correction to the bound-muon spectrum due to nuclear polarization, given by the sum of the (b) ladder, (c) crossed, and (d) seagull two-photon-exchange diagrams.}
\label{fig:TPE}
\end{center}
\end{figure}

We express the two-photon-exchange amplitude of Figure~\ref{fig:TPE} as products of leptonic $L_{\mu\nu}(q,k)$, gauge $D_{\mu\nu}(q)$, and hadronic $H_{\mu\nu}(q,-q)$ tensors. The ladder (b) and crossed (c) diagrams share this factorized structure, while the seagull diagram (d) is a contact term that does not factorize and enters through the two-photon piece $B_{\perp}$ of the hadronic tensor [see Eq.~(\ref{eq:muonic_HT})]. The amplitude reads
\begin{equation}
\iu \mathcal{M} = \int \frac{d^4q}{(2\pi)^4} \ L_{\mu\nu}(q,k) \ D^{\nu\sigma}(q) \ D^{\mu\rho}(-q) \ H_{\sigma\rho}(q,-q) \ ,
\label{eq:muonic_tpe_amplitude}
\end{equation}
where $q$ is the four-momentum of the photon loop and $k$ the muons' four-momentum.

Since muons are structureless spin-$1/2$ particles, the leptonic tensor, $L^{s}_{\mu\nu}(q,k)$, is obtained at this order in the QED expansion:
\begin{equation}
L^{s}_{\mu\nu}(q,k) =
\Biggl[- \frac{\iu e^2}{(k-q)^2 - m^2 + \iu \epsilon}  \Biggr]
\bar{u}^s(k) \ \gamma_\mu \Bigl[ (k-q)^\lambda \gamma_\lambda + m \Bigr] \gamma_\nu \ u^{s}(k) \ ,
\label{eq:muonic_leptonic_tensor}
\end{equation}
where $u^s(k)$ are the Dirac wavefunctions of the bound muon with spin $s$, while $m$ denotes the muon mass and $e$ the absolute value of its electric charge.

In both the diagrams of Figure~\ref{fig:TPE} and in Eq.~(\ref{eq:muonic_leptonic_tensor}), we set the spins of the incoming and outgoing muons to be equal. Spin-flip processes are generally associated with hyperfine-splitting corrections. The internal leptonic line has been evaluated using the free Dirac propagator, which yields the leading two-photon-exchange contribution to the Lamb shift, of order $(Z\alpha)^5$. Corrections to this approximation, in which the intermediate muon line is instead evaluated in the Coulomb field of the nucleus, are referred to as Coulomb-distortion effects and are of order $(Z\alpha)^6$; they are thus largely neglected in recent works. However, one Coulomb-distortion correction has been found to be logarithmically enhanced, of order $(Z\alpha)^6\log(Z\alpha)$, and is retained in recent calculations \cite{Fri77,Pac11,Ji18,Pac23,Lim25}. Since the Lamb shift of muonic atoms is not sensitive to the spin-spin interaction between lepton and nucleus, one can simplify the expressions by averaging the leptonic tensor in Eq.~(\ref{eq:muonic_leptonic_tensor}) over the spin.

Eq.~(\ref{eq:muonic_leptonic_tensor}) has been written in the forward limit, where there is no overall momentum transfer to the muon during the two-photon process. Furthermore, one generally neglects the three-momentum of the lepton with respect to its rest mass. This approximation is motivated by the observation that light hydrogen-like atoms are loosely bound systems, where the momentum of the bound muon scales as $\vert \mathbf{k} \vert \sim m (Z\alpha)$. For light atoms, this implies $\vert \mathbf{k} \vert \ll m$. The derivations presented here are based on the applicability of this approximation, meaning that the analysis is suited to light systems where $(Z\alpha)$ is small.

The hadronic tensor satisfies crossing symmetry, $H_{\rho\sigma}(-q,q)=H_{\sigma\rho}(q,-q)$~\cite{Ros83}. The symmetries of the amplitude in Eq.~(\ref{eq:muonic_tpe_amplitude}) can then be exploited by writing the leptonic tensor in a way that follows this symmetry as well. Adding the crossed diagram, which supplies the second time ordering with $(k+q)$ flowing through the internal line and hence the denominator $q^2+2mq_0$, and averaging over the muon spin, one obtains
\begin{equation}\resizebox{0.9\textwidth}{!}{$\displaystyle
    L^{(S)}_{\mu\nu} = - \iu 2\pi\alpha \Biggl[
    \frac{2m\delta_{\nu 0}\delta_{\mu 0} - q_\mu \delta_{\nu 0} - q_\nu  \delta_{\mu 0} + q^0 g_{\mu\nu}}{q^2-2mq^0+\iu\epsilon}
    +
    \frac{2m\delta_{\nu 0}\delta_{\mu 0} + q_\mu \delta_{\nu 0} + q_\nu \delta_{\mu 0} - q^0 g_{\mu\nu}}{q^2+2mq^0+\iu\epsilon}
    \Biggr] \ .$}
    \label{eq:muonic_symmetrized_leptonic_tensor}
\end{equation}
As a next step, we specify a gauge for the photon propagators in Eq.~(\ref{eq:muonic_tpe_amplitude}). Our calculation is performed in Coulomb gauge, $(\nabla \cdot \mathbf{A} = 0)$, which significantly simplifies the final expression for the amplitude and allows for a simpler numerical evaluation of the corrections. The photon propagator in this gauge takes the form
\begin{align}
D^{\mu\nu}(q) \ \ &= \ \ \left\{
\begin{array}{l}
D^{00}(q) =\dfrac{1}{\mathbf{q}^2} \\
D^{i0}(q) = D^{0j}(q) = 0 \\
D^{ij}(q) = \dfrac{1}{q^2+\iu\epsilon} \Bigl(\delta^{ij}-\dfrac{q^iq^j}{\mathbf{q}^2} \Bigr)
\end{array}\right. \ ,
\label{eq:muonic_coulomb_gauge_propagator}
\end{align}
where Greek letters denote Lorentz indices and Roman letters denote spatial components, namely $i,j=1,2,3$.

The simplifications obtained in Coulomb gauge become immediately clear when performing the contractions with the symmetrized leptonic tensor $L^{(S)}_{\mu\nu}$ of Eq.~(\ref{eq:muonic_symmetrized_leptonic_tensor}). Many combinations vanish and one arrives at
\begin{equation}
L^{(S)}_{\mu\nu} \ D^{\nu\sigma}(q) \ D^{\mu\rho}(-q)
=
\iu (4\pi\alpha) \frac{2m}{(q^2+\iu\epsilon)^2-4m^2q_0^2}
\Biggl[
\frac{\delta^{\sigma}_{\,0}\,\delta^{\rho}_{\,0}}{\mathbf{q}^2}
+
\frac{q_0^2}{(q^2+\iu\epsilon)^2} \ T^{\sigma\rho}
\Biggr] \ ,
\label{eq:muonic_contracted_amplitude}
\end{equation}
where the free indices $\sigma,\rho$ are those that contract with the hadronic tensor $H_{\sigma\rho}$. The first term in brackets carries $\sigma=\rho=0$ and therefore projects onto the charge-charge response, while $T^{\sigma\rho}$ is the transverse projector, non-zero only for spatial indices $\sigma=i$, $\rho=j$,
\[
T^{ij}=\delta^{ij} - \frac{q^iq^j}{\mathbf{q}^2} \ .
\]
Note that this is not to be confused with the generalized Compton tensor which appears in Section~\ref{section:superallowed}.

\subsection{The \texorpdfstring{$ \eta $}{eta} expansion and multipole sum rules}
\label{sec:muonic_eta_expansion_sum_rules}

The total amplitude follows from combining Eq.~(\ref{eq:muonic_tpe_amplitude}), Eq.~(\ref{eq:muonic_contracted_amplitude}), and the electromagnetic hadronic tensor in Eq.~(\ref{eq:em_hadronic_tensor}), leading to
\begin{equation}
\iu \mathcal{M} = \iu (4\pi\alpha)^2 \int \frac{d^4q}{(2\pi)^4}
\ \frac{2m}{(q^2+\iu\epsilon)^2-4m^2q_0^2}
\ \Biggl[
\frac{1}{\mathbf{q}^2} \ H_L
+
\frac{q_0^2}{(q^2+\iu\epsilon)^2} \ H_T
\Biggr] \ ,
\label{eq:muonic_HL_HT_amplitude}
\end{equation}
where we factored out an overall $(4\pi\alpha)$ from the definition of $H_{\rho\sigma}$ and defined the longitudinal $H_L$ and transverse $H_T$ components of the hadronic tensor as
\begin{eqnarray}
H_L &=& \sum_{N \neq 0} \lvert \bra{N} \rho_{\text{ch}}(\mathbf{q}) \ket{0}\rvert ^2
\Biggl[
\frac{1}{q_0 - \omega_N + \iu \epsilon}
-
\frac{1}{q_0 + \omega_N - \iu \epsilon}
\Biggr] \ ,
\label{eq:muonic_HL} \\
H_T &=& \sum_{N\neq 0} \lvert \bra{N} \mathbf{J}_{\perp} \ket{0} \rvert^2
\Biggl[
\frac{1}{q_0-\omega_N +\iu\epsilon}
-
\frac{1}{q_0+\omega_N-\iu\epsilon}
\Biggr]
+
B_{\perp}(\mathbf{q}) \ ,
\label{eq:muonic_HT}
\end{eqnarray}
where the symbol $(\perp)$ indicates contraction with the transverse tensor $T^{ij}$ defined after Eq.~(\ref{eq:muonic_contracted_amplitude}), while $\rho_{\text{ch}}(\mathbf{q})=J_0(\mathbf{q})$ and $\omega_N=E_N-E_0$. Equations~(\ref{eq:muonic_HL}) and~(\ref{eq:muonic_HT}) follow from the spectral form of the elastic forward tensor, Eq.~(\ref{eq:em_hadronic_tensor}), evaluated at $\ket{i}=\ket{f}=\ket{0}$, after using $E_0-E_N\pm q_0=\pm q_0\mp\omega_N$. The sums are restricted to $N\neq 0$: the elastic ground-state term is independent of the nuclear excitation spectrum and is reabsorbed into the elastic (Zemach) part of the finite-size correction, so that $H_L$ and $H_T$ collect the genuinely inelastic, polarizability response of the nucleus.

The amplitude $\iu\mathcal{M}$ acts as an effective, complex muon--nucleus interaction whose real part shifts a bound-muon level $\ket{\phi_{nL}}$ by $\Delta E_{nL}=\bra{\phi_{nL}}\mathrm{Re}(\iu\mathcal{M})\ket{\phi_{nL}}$. Since the muon three-momentum is neglected, this expectation value collapses to the forward amplitude weighted by $\phi^2(0)\equiv\lvert\phi_{nL}(0)\rvert^2$, the probability of finding the muon at the center of the nucleus. The evaluation of the amplitude is generally organized in two steps: the dominant non-relativistic corrections are first obtained from the $\eta$ expansion described in the remainder of this subsection, and the smaller relativistic corrections are subsequently added on top, as detailed below.

The non-relativistic corrections follow from the approximation ($q_0,\lvert\mathbf{q}\rvert\ll m$) of the integration kernels in the amplitude Eq.~(\ref{eq:muonic_HL_HT_amplitude}). In this limit the transverse kernel carries an explicit factor $q_0^2$ and is therefore suppressed, so that the transverse response $H_T$ vanishes identically in the leading non-relativistic term and the whole $\eta$ expansion is governed by the longitudinal (charge) response $H_L$ \cite{Fri97a,Fri13,Pac15a}. The kernels are expanded in terms of the small parameter $\eta=\sqrt{2m_r\omega_N}\,\lvert\mathbf{R}-\mathbf{R}'\rvert$, with $\mathbf{R},\mathbf{R}'$ the positions of the struck protons; its expectation value scales as $\sqrt{m_r/m_p}\ll1$, taking values $\eta\approx0.15$ and $0.35$ for $\mu^4\mathrm{He}^+$ and $\mu^3\mathrm{He}^+$, respectively~\cite{Lim22}. Performing the $q_0$ and $\lvert\mathbf{q}\rvert$ integrations and expanding order by order in $\eta$ then produces a sequence of energy-weighted multipole sum rules, which can efficiently be evaluated with the Lanczos techniques of Section~\ref{Section: 3.2}, see also Ref.~\cite{PRC.89.064317.2014}. The leading term is the electric-dipole polarizability contribution,
\begin{equation}
\delta^{(0)}_{D1} = -\frac{16\pi^2}{9}(Z\alpha)^2\,\phi^2(0)\,\frac{1}{2J_0+1}\sum_{N\neq 0}\sqrt{\frac{2m_r}{\omega_N}}\,\lvert\langle N\lVert \hat{D}_1\rVert 0\rangle\rvert^2 \ ,
\label{eq:muonic_dipole_polarizability}
\end{equation}
where $J_0$ is the ground-state spin and $\langle N\lVert \hat{D}_1\rVert 0\rangle$ are the reduced matrix elements of the dipole operator $\hat{D}_1$, i.e., the long-wavelength limit of the $J=1$ Coulomb multipole $M_{1M}$ of Section~\ref{sec:multipoles_current_operators}. Equation~(\ref{eq:muonic_dipole_polarizability}) shows that the dominant nuclear-structure correction is governed by the $\omega^{-1/2}$-weighted moment of the dipole strength, closely related to the static dipole polarizability $\alpha_D$.

At sub-leading order in $\eta$ two terms appear, denoted $\delta^{(1)}_{Z3}$ and $\delta^{(1)}_{R3}$ in Ref.~\cite{Ji18}. The first reproduces, with opposite sign, the elastic Zemach (Friar) contribution
\begin{equation}
\delta_{\text{Zem}} = -\frac{m_r^4}{24}(Z\alpha)^5\int\!\!\int d^3R\,d^3R'\,\lvert\mathbf{R}-\mathbf{R}'\rvert^3\,\rho_{\text{ch}}(\mathbf{R})\,\rho_{\text{ch}}(\mathbf{R}') \ ,
\label{eq:muonic_zemach}
\end{equation}
proportional to the third (Friar) moment of the charge density $\rho_{\text{ch}}$: in the point-nucleon limit $\delta^{(1)}_{Z3}=-\delta_{\text{Zem}}$, so that the elastic piece cancels out of the total two-photon-exchange correction. Its inelastic counterpart $\delta^{(1)}_{R3}$, built on the proton-proton correlation function, enjoys no such cancellation: it vanishes for the hydrogen isotopes, which contain a single proton, but for $Z>1$ it survives in full and is by no means small. In $\mu^3\mathrm{He}^+$, for instance, $\delta^{(1)}_{Z3}$ and $\delta^{(1)}_{R3}$ amount to $+8.3$ and $-8.7$~meV, respectively, each individually exceeding in magnitude the leading dipole term $\delta^{(0)}_{D1}\simeq-6.6$~meV~\cite{Ji18}. At the following order three further sum rules arise -- an electric monopole, an electric quadrupole, and a dipole-like ($D_1$-$D_3$) (see Ref.~\cite{Ji18} for the definition of the $D_3$ operator) interference response -- which complete the non-relativistic $\eta$-expansion at the accuracy required for the $(Z\alpha)^5$ correction.

Two additional classes of corrections are retained at this precision. The Coulomb-distortion effect introduced in Section~\ref{sec:muonic_tpe_amplitude} generates a further electric-dipole sum rule carrying an unusual logarithmic energy weight, $\delta^{(0)}_C\propto(Z\alpha)^6\log(Z\alpha)$, which is numerically non-negligible. Secondly, relativistic corrections are added on top of the non-relativistic $\eta$ expansion. They are obtained by retaining the full relativistic kinematics of the integration kernels in the amplitude Eq.~(\ref{eq:muonic_HL_HT_amplitude}) and subtracting the leading non-relativistic contribution of Eq.~(\ref{eq:muonic_dipole_polarizability}) already included above. As opposed to the non-relativistic terms which are purely longitudinal, the relativistic corrections are both longitudinal and transverse. Due to their suppression, these terms are evaluated only in the leading dipole approximation. The foundational work evaluating these non-relativistic and relativistic expansions for the nuclear polarizability in light muonic atoms was carried out in a series of studies~\cite{Fri77,Fri13,Pac07,Pac11,Ji13,Nev16,Pac15a,Pac18,Ji18}.
\subsection{Recent \textit{ab initio} calculations}
\label{sec:muonic_ab_initio_bayesian}

The previous section focused on the nuclear structure corrections calculated at the level of the TPE term, namely up to order $(Z\alpha)^5$. The complete nuclear-structure correction can be schematically organized as
\begin{equation}
\delta_{\text{NS},\mu} = \delta_{\text{TPE}} + \delta_{\text{3PE}} + \delta_{\text{EVP}} + \delta_{\text{MSEVP}} + \cdots \ ,
\label{eq:muonic_dNS_expansion}
\end{equation}
where the dominant two-photon-exchange term splits into a nuclear and a single-nucleon part, $\delta_{\text{TPE}}=\delta^A_{\text{TPE}}+\delta^N_{\text{TPE}}$, the former computed \textit{ab initio}; the remaining terms $\delta_{\text{3PE}}$, $\delta_{\text{EVP}}$, and $\delta_{\text{MSEVP}}$ are the three-photon-exchange~\cite{Pac18}, the electron vacuum-polarization, and the muon self-energy and vacuum-polarization corrections~\cite{Pac23}, respectively.

Modern calculations evaluate the nuclear part $\delta^A_{\text{TPE}}$ from the sum rules of Section~\ref{sec:muonic_eta_expansion_sum_rules} with the many-body machinery of Section~\ref{sec:modern_ab_initio_methods}. The ground state $\ket{0}$ follows from a $\chi$EFT Hamiltonian solved with the effective-interaction hyperspherical-harmonics (EIHH) method, which reaches sub-percent accuracy for the three- and four-body nuclei~\cite{Lim21,Ji18,Lim25}, while the energy-weighted response is obtained with the Lanczos-based methods of Section~\ref{Section: 3.2}, bypassing the explicit construction of the excited states $\ket{N}$ -- the same response machinery later applied to the $\gamma W$-box amplitude in Section~\ref{section:superallowed}. The $\chi$EFT truncation error is propagated through the Bayesian analysis of Section~\ref{Section: 3.3} [Eq.~(\ref{eq:uq_eft_expansion})]. Applied to muonic atoms, this replaces the earlier practice of estimating the uncertainty by comparing a $\chi$EFT prediction with a phenomenological potential, placing the error bars on a firm statistical footing~\cite{Lim22,Lim25,Ach23}.

State-of-the-art results at N$^3$LO for muonic helium are summarized in Table~\ref{tab:muonic_helium_ns}. They give $\delta^A_{\text{TPE}}=-14.868(364)$~meV in $\mu^3\mathrm{He}^+$ and $-8.751(322)$~meV in $\mu^4\mathrm{He}^+$~\cite{Lim25}, to be compared with the previous values $-14.72(31)$ and $-8.49(39)$~meV~\cite{Ji13,Ji18}. The Bayesian analysis increases the $\mu^3\mathrm{He}^+$ uncertainty by $17\%$ and reduces the $\mu^4\mathrm{He}^+$ one by $17\%$, placing both on a common statistical basis. The associated dipole polarizabilities, ${\alpha}_{\text{D}}=3.514(68)$ and $1.909(96)$~fm$^3$ for $^3$He and $^4$He, agree with earlier determinations~\cite{Pac07}. Combined with the remaining terms of Eq.~(\ref{eq:muonic_dNS_expansion}) from \cite{Pac23} and the measured Lamb shifts \cite{Sch25,Kra21}, these corrections yield the charge radii listed in Table~\ref{tab:muonic_helium_ns}.

\begin{table}[h]
\centering
\caption{Nuclear-structure corrections in muonic helium at N$^3$LO and the extracted charge radii, from Ref.~\cite{Lim25}. Uncertainties are obtained from a Bayesian analysis of the $\chi$EFT truncation.}
\label{tab:muonic_helium_ns}
\begin{tabular}{lccc}
\toprule
 & $\delta^A_{\text{TPE}}$ (meV) & $\delta_{\text{NS},\mu}$ (meV) & $r_c$ (fm) \\
\midrule
$^3$He & $-14.868(364)$ & $-15.644(427)$ & $1.9704(11)$ \\
$^4$He & $-8.751(322)$  & $-9.541(368)$  & $1.6793(10)$ \\
\bottomrule
\end{tabular}
\end{table}

\subsection{Impact on the helium isotope-shift puzzle}
\label{sec:muonic_helium_isotope_shift_puzzle}

A particularly sensitive observable is the difference of the squared charge radii of the two helium isotopes, $\delta r^2 = r_c^2(^3\mathrm{He}) - r_c^2(^4\mathrm{He})$. Because the dominant theoretical uncertainties are strongly correlated between $^3$He and $^4$He, $\delta r^2$ can be determined more precisely than either radius individually~\cite{Lim25}. It is moreover accessible by two independent routes; from the absolute radii measured in muonic helium through the Lamb shift, and from the isotope shift of the $2\,^3S\to2\,^1S$ transition in electronic helium.

Using the muonic-helium nuclear-structure corrections of Section~\ref{sec:muonic_ab_initio_bayesian}, the muonic route gives $\delta r^2|_\mu = 1.0626(29)$~fm$^2$~\cite{Lim25}, a $6\%$ improvement of the uncertainty over the value $1.0636(6)_{\text{expt}}(30)_{\text{theo}}$~fm$^2$ extracted directly from the CREMA measurements~\cite{Sch25}, while the ordinary-atom route gives $\delta r^2|_e = 1.0758(15)$~fm$^2$~\cite{Lim25}, basically matching the value $1.0757(15)$~fm$^2$ of Ref.~\cite{Van23} obtained using the previous theoretical calculation for the nuclear structure effects. At the time of the analysis of Ref.~\cite{Lim25} the two routes disagreed at the $4\sigma$ level; far from resolving the tension, the improved nuclear-structure input increased the earlier $3.6\sigma$ discrepancy~\cite{Sch25}. Crucially, the Bayesian framework showed that the shift needed to reconcile the two routes was excluded on the nuclear-structure side at the $95\%$ confidence level~\cite{Lim25}, exonerating the nuclear corrections and pointing instead to experimental systematics or neglected atomic-theory contributions.

The resolution came from the atomic theory of ordinary helium. Qi and collaborators~\cite{Qi25} identified a previously neglected hyperfine mixing of the $2\,^1S$ and $2\,^3S$ levels of $^3$He with higher $S$ states ($n>2$), which shifts the $2\,^3S\to2\,^1S$ isotope shift and lowers the ordinary-atom value to $\delta r^2|_e = 1.0693(15)$~fm$^2$, reducing the tension with respect to Ref.~\cite{Sch25} to $1.7\sigma$. Pachucki and collaborators subsequently derived the complete second-order hyperfine correction~\cite{Pac24}, obtaining $\delta r^2|_e = 1.0678(7)$~fm$^2$, which agrees with the muonic value of Ref.~\cite{Sch25} at the $1.3\sigma$ level and with the improved determination of Ref.~\cite{Lim25} at $1.7\sigma$. The resulting consistency of the two routes validates the extraction of absolute charge radii from muonic atoms and supports its extension to the heavier systems.

\subsection{Heavy muonic atoms and highly charged ions}
\label{sec:heavy_muonic_atoms}

While the \textit{ab-initio} methods combined with $\chi$EFT discussed above provide rigorous uncertainty quantification for the lightest nuclei ($A\leq4$), they are not feasible for the heavy systems probed by muonic X-ray spectroscopy -- which accesses nuclear charge radii through the energies of the $np\to1s$ transitions emitted during the muonic-atom cascade, and has recently experienced a strong revival. On the experimental side, the QUARTET collaboration at the Paul Scherrer Institute aims at measuring the absolute charge radii of light nuclei ($3\leq Z\leq 10$, from lithium to neon) with up to an order-of-magnitude improved accuracy by detecting the low-energy muonic X~rays with metallic magnetic calorimeters~\cite{Oha24,Ung24}. In parallel, the muX project has developed a repeated muon-transfer technique in a high-pressure hydrogen--deuterium gas cell that enables muonic-atom spectroscopy with microgram amounts of target material, giving access to rare and radioactive species such as $^{226}$Ra and $^{248}$Cm~\cite{Wau21,Ada23}. This program has recently been extended to the medium-mass region, where a high-statistics measurement of muonic $^{35,37}$Cl with a large-scale germanium detector array determined the charge radii of the stable chlorine isotopes with an order-of-magnitude improvement in precision~\cite{Bey25}. The chlorine measurement was accompanied by a dedicated theory framework for medium-mass muonic atoms~\cite{Rat25}. Across this entire program, the nuclear polarizability correction remains the most challenging and uncertain theoretical input, and its evaluation has consequently attracted renewed attention~\cite{Lim19,Rat25,Gor25,Val22,Val24}.

This scrutiny is driven by a long-standing puzzle: the $nP_{1/2}$--$nP_{3/2}$ fine-structure splittings measured in heavy muonic atoms such as $\mu$-$^{90}\mathrm{Zr}$, $\mu$-$^{120}\mathrm{Sn}$ and $\mu$-$^{208}\mathrm{Pb}$ show persistent anomalies between theory and experiment that have traditionally been ascribed to the nuclear polarizabilities~\cite{Val22}. Because an \textit{ab initio} description of heavy nuclear spectra is out of reach, the nuclear excitation spectrum entering the correction is instead computed within the random-phase approximation (RPA) on top of a self-consistent Hartree-Fock mean field built from an effective Skyrme interaction~\cite{Val22}. The muon-nucleus interaction is treated in a fully field-theoretical framework; the same ladder, crossed and seagull diagrams of Figure~\ref{fig:TPE} reappear~\cite{Val22,Val24}.

A systematic study across nine Skyrme parametrizations indicated that the resulting nuclear-model uncertainty on the nuclear polarizability correction is much smaller than the persistent theory--experiment discrepancies, suggesting that the nuclear polarizability is \emph{not} the origin of the fine-structure anomalies, which must instead be sought in refined QED or recoil corrections~\cite{Val22}. This conclusion has recently been substantiated: a re-evaluation of muonic $^{208}$Pb combining state-of-the-art QED with modern nuclear polarizability models closed the long-standing fine-structure-anomaly gap~\cite{Sun25}, while an all-order treatment of the relativistic-recoil effect resolved the analogous puzzle in muonic $^{90}$Zr~\cite{Bey25b}.

\section{Recent Progress in Superallowed Nuclear Beta Decays}\label{section:superallowed}

\begin{figure}[t]
    \begin{center}
        \includegraphics[width=\textwidth]%
        {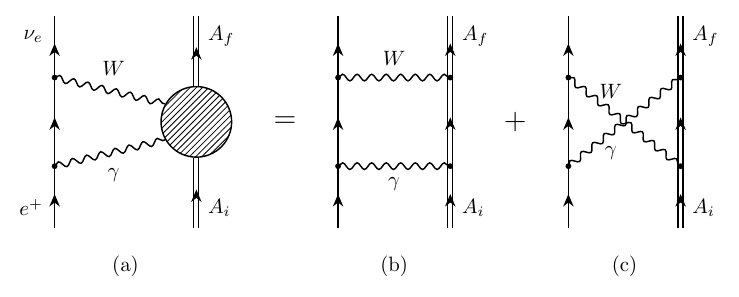}
    \end{center}
    \caption{\label{section:superallowed:figure:treelevel_and_radiativecorrection}
    Pictured is the contribution from a higher-order, virtual electroweak box diagram to the beta decay process of $ \Psi_{i} \rightarrow \Psi_{f} + \nu_{e} + e^{+} $, known as the $ \gamma W $ box.}
\end{figure}

Precision beta decay provides one of the cleanest low-energy probes of the weak-charged-current interaction~\cite{PRC.102.045501.2020, P.4.397467.2021}. Nuclear Fermi transitions, a special type of weak transition in nuclei, are decays of the kind $ ( J^{\pi}, \, T \, ) = ( 0^{+}, \, 1) \rightarrow ( 0^{+}, \, 1 ) $ for which the leading amplitude is heavily constrained by the spin-isospin symmetries of the external states. These transitions give clean access to the Cabibbo-Kobayashi-Maskawa (CKM) up-down-quark-mixing element $ V_{ud} $ and, hence, to the top-row unitarity test of the SM. At the quark level, the weak-charged current is given by the Lagrangian
\begin{align}
    \mathcal{L}_{ \mathrm{WCC} } =  - \frac{g}{ 2 \sqrt{2} } \
    \begin{pmatrix}
        \, \bar{u}_{L} \! & \! \bar{c}_{L} \! & \! \bar{t}_{L} \,
    \end{pmatrix}
    \; \gamma^\mu \, W^+_\mu \, V_{ \mathrm{CKM} } \;
    \begin{pmatrix}
        \, d_{L} \\ s_{L} \\ b_{L} \,
    \end{pmatrix} \
    + \ \mathrm{h.c.}
    \quad ,
\end{align}
where $ e = g \sin( \theta_{\mathrm{W}} ) $ defines the weak coupling constant and the CKM matrix enters as a $ 3 \times 3 $ unitary matrix in flavour space~\cite{PRL.10.531533.1963, PTP.49.652657.1973} with the top-row unitarity condition
\begin{align}
    \vert V_{ud} \vert^{2} + \vert V_{us} \vert^{2} + \vert V_{ub} \vert^{2} = 1
    \quad .
\end{align}
At the precision of $ \mathcal{O}(10^{-4})$ required for modern tests of CKM matrix unitarity, the tree-level amplitude is not sufficient and electroweak radiative corrections (EWRC) must be explicitly included in extraction of $V_{ud}$; a relevant process is the $ \gamma W $ box whose basic physics is illustrated in Figure~\ref{section:superallowed:figure:treelevel_and_radiativecorrection}. Evaluation of such radiative corrections in nuclear theory have become the defining challenge in precision extractions of $ V_{ud} $. A general current-algebra framework developed some time ago by A. Sirlin~\cite{RMP.50.573605.1978} forms the basis of the analysis presented in the subsequent sections.

We will assume isospin symmetry in the following relativistic derivations, however, it should be noted that isospin-breaking effects are included in all nuclear-structure-dependent pieces of the calculation. This is well-enough justified as superallowed transitions occur between nearly-degenerate isospin analog states. Further note that the discussion is tailored towards nuclear decays.


\subsection{Extraction of \texorpdfstring{$V_{ud}$}{Vud} from Superallowed Decays}\label{section:superallowed:section:physical_context}

For a given nuclear Fermi transition $ \Psi_{i}(p) \longrightarrow \Psi_{f}(p') + e^{+}(p_e) + \nu_{e}( p_{\nu} ) $, the tree-level amplitude between relativistic lepton-nucleus, plane-wave states is
\begin{align}
    i M_{0} \big( p_{f}, p_{i}, k_{f}, k_{i} \big)
    = - i \frac{ G_{F} V_{ud}^{*} }{ \sqrt{2} } \ L_{ \lambda } ( k_{f}, k_{i} ) \ F^{\lambda}( p_{f}, p_{i} )
    \quad ,
\end{align}
where $ F^{\lambda}(p',p) $ is the amplitude of the charged-weak current with external $ J^{\pi} = 0^{+} $ and $ T=1 $ nuclear states, $ L_{ \lambda } $ is the leptonic amplitude, and $ G_{F} / \sqrt{2} = g^{2} / 8 M_{W}^{2} $ is the Fermi constant. They are respectively given by
\begin{gather}
    F^{\lambda}( p_{f}, p_{i} )
    = \planewavebra{ f }{ \mathbf{p}_{f} } \, J_{ \mathrm{W} }^{ \dagger \, \lambda } ( q ) \, \planewaveket{ i }{ \mathbf{p}_{i} }
    = f_{+} (t) \, ( p_{f} + p_{i} )^{\lambda} + f_{-} (t) \, ( p_{f} - p_{i} )^{\lambda}
    \quad , \nonumber \\[0.33cm]
    L^{\lambda} = \bar{u}_e \gamma^{\lambda} ( 1 - \gamma_5 ) v_{\nu}
    \quad , \nonumber
\end{gather}
with $ t = ( p_{f} - p_{i} )^{2} $ and the former written in the isospin symmetric limit with spinless external states (exact for superallowed transitions). These form factors are normalized such that $f_{+}(0) = \sqrt{2}$ and $f_{-}(0) = 0$ in $ T=1 $ systems. At leading order in $ \vert \mathbf{p} \vert  / m $, the axial-vector piece of the current does not contribute and the conserved-vector-current hypothesis implies that the vector matrix element is fixed by isospin symmetry to be $ M_{F0} = \sqrt{2} $.

Then, for superallowed transitions, one may write down the differential rate in the point-like nucleus limit as
\begin{align}
    d\Gamma = dE_{e} \ \frac{ G_{F}^2 \, g_{v}^{2} \, m_{e}^{5} \, \big\vert M_{F0} \big\vert^{2} \big\vert V_{ud} \big\vert^{2} }{ 2 \pi^{3} } \
    \frac{1}{ m_{e}^{5} } \, \vert \vec{p}_{e} \vert \, E_{e} \, ( E_{0} - E_{e} )^{2} \, F \big( Z_{f}, E_{e} \big)
    \quad ,
\end{align}
where $ F( Z, E_{e} ) $ is the Fermi function~\cite{ZP.88.161177.1934, PRD.109.056006.2024, PRL.133.021803.2024}. Many corrections arise in this expression as non-point-like distributions and higher-order processes for the nucleus are considered, e.g., see Section 3.1 of Ref.~\cite{ARNPS.74.2347.2024} for a summary. After integration over the positron phase space to produce the dimensional statistical phase-space factor $ f $ and multiplication by the half life of the decay process $ t = \ln 2 / \Gamma $, we have
\begin{align}
    \big\vert V_{ ud } \big\vert^{2} = \frac{ K }{ G_{F}^{2} \, g_{v}^{2} \, \vert M_{F0} \vert^{2} ft } + \text{higher-order corrections}
    \quad ,
\end{align}
where $ K $ is a transition-independent constant. The promise of these decays is simple: any measurement of a single or set of the available superallowed transitions should lead to a consistent extraction of $ V_{ud} $. However, this is hardly the case at the current precision level due to the aforementioned requirement of corrections to the decay rate.

As shown by Sirlin, use of the conserved-vector-current hypothesis and electromagnetic Ward identities allows one to express the combination of vertex corrections and radiative diagrams at $ \mathcal{O}( G_{F} \, \alpha ) $ into a set of structure-independent universal logarithms, a long-distance outer correction, and a hadronic-structure-sensitive inner correction. Schematically, the correction to the amplitude is then given in terms $ \mathcal{M}_{0} $ as
\begin{align}
    \delta \mathcal{M}
    =  \mathcal{M}_{0} \, \frac{\alpha}{2\pi} [ \ \text{universal logs} + \delta_{ \mathrm{outer} } + \delta_{ \mathrm{inner} } \ ]
    + \mathcal{O}( G_{F} \, \alpha^2 )
    \quad .
\end{align}
The dominant part of the \textit{outer} correction $\delta_R'$ accounts for infrared QED effects, e.g., bremsstrahlung photon emission, as well as Coulomb distortion of the positron wave function~\cite{PRJA.164.17671775.1967, PRL.57.19941997.1986, PRD.35.34203422.1987, PRD.35.34233427.1987, PRD.70.093006.2004}; typically quoted to order $ Z \alpha^{3} $, the outer correction has seen increased scrutiny in recent years which call into question certain results from historical works~\cite{PRL.133.021803.2024, PRD.109.056006.2024, PLB.868.139678.2025, arXiv:2511.05446}. The short-range analogue known as the \textit{inner} correction $ \Delta^{V}_{R} $ is universal to all superallowed transitions and comprises several effects, the most challenging of which to compute is the hadronic-structure-dependent $ \gamma W $ box which dominates the uncertainty; theoretical errors were greatly improved in the last eight years~\cite{PRL.121.241804.2018, PRD.100.013001.2019, PRD.101.111301.2020, PRD.104.033003.2021, PRD.100.073008.2019} and led directly to the observed tension with unitarity. In addition to the QED and intra-nucleon-structure effects, $ \beta $ decay in composite hadrons like nuclei begets additional structure corrections, namely, $ \delta_{\mathrm{C}} $ and $ \delta_{ \mathrm{NS} } $. The former is known as the isospin symmetry breaking correction and it parameterizes the renormalization of the Fermi amplitude due to isospin-non-conserving interactions, i.e., the strong and Coulomb interactions. The latter parameterizes the modification of the free-nucleon $ \gamma W $ box due to the nuclear medium, and one may approximately separate the nucleon- and nuclear-level effects as
\begin{align}
    1 + \delta_{\mathrm{inner}}
    = 1 + \Delta^{U}_{R} + 2 \square_{\gamma W}^{n} + 2 ( \mathrm{Re} \, \square_{\gamma W}^{\mathrm{nuc.}} - \square_{\gamma W}^{n} )
    \approx ( 1 + \Delta^{V}_{R} ) ( 1 + \delta_{\mathrm{NS}} )
    \quad ,
\end{align}
where $ \Delta^{U}_{R} $ (and consequently $ \Delta^{V}_{R} $) is a piece universal to all superallowed decays, $ \square_{ \gamma W }^{n} $ is the free-nucleon box contribution, and $ \square_{\gamma W}^{\mathrm{nuc.}} $ is the nuclear box contribution in a given transition~\cite{ARNPS.74.2347.2024}. After the appropriate SM corrections are applied we have
\begin{equation}\label{sec:prelim:Vud_superallowed}
    \vert V_{ud} \vert^2 = \frac{ K }{ G_{F}^{2} \, g_{v}^{2} \, \vert M_{F0} \vert^{2} \mathcal{F}t \, ( 1 + \Delta^{V}_{R} ) }
    \qquad \quad
    \mathcal{F}t = ft ( 1 + \delta_{R}' ) ( 1 - \delta_{\mathrm{C}} + \delta_{\mathrm{NS}} )
    \quad ,
\end{equation}
and different nuclei should yield a common corrected $ \mathcal{F}t $ value to the present desired precision level, and hence a consistent value of $ V_{ud} $.

\begin{figure}[t!]
    \center\includegraphics[width=0.95\textwidth]%
    {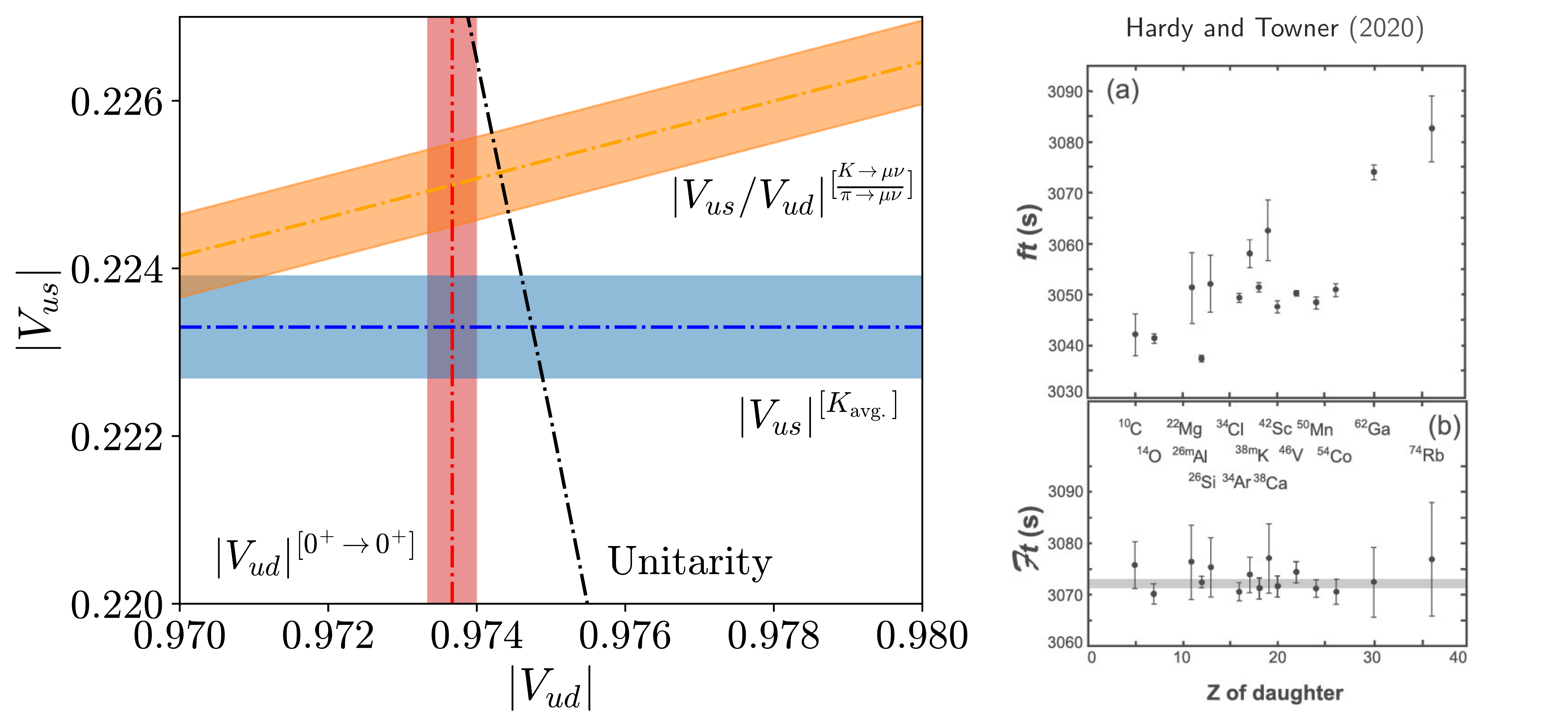}
    \caption{\label{section:introduction:figure:CKM_unitarity_and_HT_Ft_values}
    \textbf{(Left)}
    The modern CKM unitarity landscape with values for $ V_{ud} $, $ V_{us} $ and the ratio $ V_{ud} / V_{us} $ annotated with their corresponding experimental extractions~\cite{Nav24}. The dashdot lines and bands respectively represent the central value and uncertainty in the extraction of a given top-row element of the CKM matrix. The black line indicates the exact unitarity constraint. Note that the contribution of $ V_{ub} $ is $ \mathcal{O}(10^{-5}) $ and is thus negligible at this precision.
    \textbf{(Right)} Figures taken directly from Ref.~\cite{PRC.102.045501.2020} with evaluations for a range of nuclei from $ {}^{10}\mathrm{C} $ to $ {}^{74}\mathrm{Rb} $ of \textbf{(a)} $ ft $ values as extracted from experiment \textbf{(b)} corrected $ \mathcal{F}t $ values with higher-order SM processes included.
    }
\end{figure}

This is precisely the comparison shown on the right-hand-side of Figure~\ref{section:introduction:figure:CKM_unitarity_and_HT_Ft_values}, with the sub-figures taken from Ref.~\cite{PRC.102.045501.2020}. The issue is severe as the uncorrected $ ft $ values, shown in the upper plot \textbf{(a)}, are not transition independent at the required precision level. This is in contrast to the corrected $ \mathcal{F}t $ values, shown in the lower plot \textbf{(b)}, which encode the electroweak radiative, recoil, and nuclear-structure corrections. Clearly, the $ \mathcal{O}(10^{-4}) $ effects do much to restore the expected transition-independent behaviour. Historically speaking, the only approach to successfully produce a quasi-alignment of the nucleus-independent $ \mathcal{F}t $ values across the chart is that shown in Figure~\ref{section:introduction:figure:CKM_unitarity_and_HT_Ft_values}, and it is based upon shell-model-type calculations with Woods-Saxon potentials (see Ref.~\cite{PRC.102.045501.2020} for the latest review and references therein). Alternative approaches in literature have not been able to produce the same agreement obtained by Hardy and Towner, for example, Refs.~\cite{PRL.106.132502.2011, PRL.62.866869.1989}. Notwithstanding early efforts on the front of $ \delta_{ \mathrm{C} }$~\cite{PRC.57.31193128.1998, PRC.66.024314.2002} and prior to the last few years, in the last two decades the \textit{ab-initio} nuclear theory community has not pushed hard to reduce the theoretical uncertainties despite some criticism of their estimates~\cite{PRC.78.035501.2008, PRC.80.064319.2009, NPA.1027.122521.2022}.

Herein, we are primarily concerned with the $ \gamma W $ box process which gives rise to both $ \Delta^{V}_{R} $ and $ \delta_{ \mathrm{NS} } $. The present unitarity test, and any purported claim of beyond SM physics in superallowed decays, is sensitive to such electroweak effects at the $ \mathcal{O}(10^{-4}) $ level. Modern analyses report a two to three sigma tension between CKM unitarity and experiment~\cite{EPJC.80.2.2020, PRL.121.241804.2018, PRD.100.013001.2019, PRL.123.042503.2019, PLB.838.137748.2023, PRR.7.L042002.2025}, as shown in the up-to-date plot of the quoted PDG 2024 values for $ \vert V_{ud} \vert $ and $ \vert V_{us} \vert $ on the left-hand-side of Figure~\ref{section:introduction:figure:CKM_unitarity_and_HT_Ft_values}. Importantly, from Ref.~\cite{PRC.91.025501.2015} to Ref.~\cite{PRC.102.045501.2020}, the overall uncertainty inflated by a factor of $ 2\mathrm{x} $ after recognition that the reported nuclear-structure uncertainties were far less under control than previously thought. \textit{Ad hoc} modelling of the quasi-elastic nuclear response to the external probes via Fermi gas theory elucidated the unknown sensitivity to the nuclear spectrum. As a result, $ \delta_{\mathrm{NS}} $ became the dominant part of the uncertainty budget to $ V_{ud} $ extraction~\cite{PRC.107.035503.2023, ARNPS.74.2347.2024}, making its evaluation from \textit{ab-initio} theory a priority of the nuclear theory community~\cite{OSTI.2023}.


\subsection{The \texorpdfstring{$ \gamma W $}{gamma-W} Box Nuclear-Structure-Dependent Radiative Correction}\label{section:superallowed:section:radiative_corrections}

Thus came a flurry of works on the nuclear-structure-dependent corrections to superallowed decays. For $ \delta_{\mathrm{NS}} $, a more rigorous calculation within the current algebra formalism managed to improve the uncertainty in the $ \carbonfermidecay $ by a factor of $ 1.6\mathrm{x} $~\cite{PRL.134.012501.2025}, while a novel effective field theory (EFT) approach which dramatically simplifies the nuclear many-body calculation was implemented for the $ \carbonfermidecay $~\cite{PRC.114.015501.2026} and $ \oxygenfermidecay $~\cite{PRL.133.211801.2024, PRC.110.055502.2024} decays, albeit with a dependence on two new low-energy constants for beta decay. Moreover, a \textit{third way} based on the manifestly covariant construction of current-current correlators in point-like effective theory has been developed~\cite{PRD.113.096011.2026, arXiv:2505.05449} which systematically organizes all corrections at one loop and beyond, and may thus be mechanically advantageous over the traditional formalism; such a theory may also be incorporated as an additional step in the tower of EFTs applied in Refs.~\cite{PRC.114.015501.2026, PRL.133.211801.2024, PRC.110.055502.2024}. Furthermore, a new perturbative approach has been proposed for the isospin-symmetry-breaking correction~\cite{PRC.109.044302.2024}, while an independent attempt at extracting $ \delta_{\mathrm{C}} $ from the Fermi matrix element in Quantum Monte Carlo saw strong dependence on the nuclear Hamiltonian~\cite{arXiv:2605.14006}. While tremendous progress has been made, these works are not the definitive statement on superallowed decays and there is much to be done in the coming years. We now discuss the current algebra formalism utilized in Ref.~\cite{PRL.134.012501.2025}, which presently provides the most precise determination of $ \delta_{\mathrm{NS}} $ in the $ \carbonfermidecay $ transition.

Applying Sirlin's representation~\cite{P.4.397467.2021, RMP.50.573605.1978} to study the one-loop radiative corrections to hadronic decays yields a number of diagrams at $ \mathcal{O}( G_{F} \, \alpha ) $, the most important of which we have identified in the previous section to be the $ \gamma W $ box diagram. This correction is the only one which truly probes the non-perturbative structure of the external hadronic states. The correction to the amplitude may be split as $ \delta \mathcal{M}_{ \gamma W } = \delta \mathcal{M}_{ \gamma W }^{a} + \delta \mathcal{M}_{ \gamma W }^{b} $ with the first term in partial cancellation with the hadronic vertex correction, leaving a residual integral negligible in the Fermi amplitude when taken in the forward limit. $ \delta \mathcal{M}_{ \gamma W }^{b} $ is then the only piece of the $ \gamma W $ diagram which remains to be evaluated (see Eq.~(30) of Ref.~\cite{P.4.397467.2021}) and it is given by
\begin{equation}
    \delta \mathcal{M}_{ \gamma W }^{b}
    = - i \frac{ e^2 }{ \sqrt{2} } G_F \, L_{\lambda} \int \frac{ d^{4}q }{ (2\pi)^4 } \, \frac{ M_W^2 }{ M_W^2 - q^2 } \,
    \frac{1}{ q^2 + i\epsilon_{2} } \, \frac{ \epsilon^{ \mu \nu \alpha \lambda } q_{\alpha} }{ ( p_e - q )^2 - m_e^2 + i\epsilon_{1} } \,
    T_{ \mu \nu }
    \quad ,
\end{equation}
where the nuclear-structure-dependent piece is the generalized Compton tensor $ T_{\mu \nu} $. Its operator content is that of the generalized hadronic tensor in Eq.~(\ref{eq:hadronic_tensor_definition_coordinate}), with one electromagnetic and one charged-weak current, and whose spectral representation is given in Eq.~\ref{section:hadronictensor:equation:emw_explicit}. Note that, following the conventions in $\beta$-decay literature~\cite{RMP.50.573605.1978, P.4.397467.2021, PRC.107.035503.2023}, the Compton tensor $ T_{\mu \nu} $ is defined with relativistically normalized external nuclear states, i.e., $ \langle f ; \mathbf{p}_{f} \vert i ; \mathbf{p}_{i} \rangle = 2 E_{ \mathbf{p}_{i} } \, (2\pi)^{3} \, \delta^{(3)} ( \mathbf{p}_{f} - \mathbf{p}_{i} ) $, in contrast to the unit-normalized, non-relativistic states used throughout Section~\ref{sec:common_theory}. The leading order matching between the non-relativistic and relativistic amplitudes is given below.

First, via separation of the weak current as $ J_{W} = J_{V} - J_{A} $, we may further decompose this contribution into vector and axial-vector pieces such that
\begin{equation}
    \mathcal{M}_{ \gamma W }^{b} = \mathcal{M}_{ \gamma W }^{b, VV} + \mathcal{M}_{ \gamma W }^{b, VA}
    \quad ,
\end{equation}
for which the vector-vector piece vanishes in the forward limit and for exact isospin symmetry since $ T^{\mu \nu} $ only depends on two external momenta; the only Lorentz covariant structures which can be built are $ g_{\mu\nu} $, $ p_\mu p_\nu $, $ p_\mu q_\nu $, $ p_\nu q_\mu $, and $ q_\mu q_\nu $, all of which vanish under contraction with $ \epsilon^{\mu\nu\alpha\lambda} q_\alpha $. Then, under this set of approximations, all that remains is the vector-axial-vector contribution which can be shown to factorize as
\begin{equation}
    i \mathcal{M}_{ \gamma W }^{b, VA} = i \mathcal{M}_{0} \times \big\langle \, \mathrm{Re} \, \square_{ \gamma W }^{ b }(E_{e}) \, \big\rangle
    \quad ,
\end{equation}
where $ \mathcal{M}_{0} $ is the tree-level amplitude and $ \langle \, \mathrm{Re} \, \square_{ \gamma W }^{b, \mathrm{nuc.} } (E_{e}) \, \rangle $ is the result of averaging the real part of the nuclear $ \gamma W $ box function over the positron phase space. In the isospin-symmetric forward limit, with $M_i=M_f=M$ and $p_i=p_f=p=(M,\mathbf{0})$, the contribution of the vector-axial-vector $\gamma W$ box may be written as~\cite{PRC.107.035503.2023}
\begin{align}\label{section:superallowed:equation:gW_box}
    \Box_{\gamma W}^{b}(E_e)
    &=
    e^2 \int \frac{d^4q}{(2\pi)^4} \frac{M_W^2}{M_W^2-q^2} \frac{1}{q^2+i\varepsilon_1} \frac{1}{(p_e-q)^2-m_e^2+i\varepsilon_2}
    \nonumber \\
    &\quad \times \frac{|\mathbf{q}|\big(|\mathbf{q}|-\nu\beta^{-1}x\big)}{\nu} \frac{T_3(\nu,\mathbf{q})}{M f_+}
    \quad ,
\end{align}
where $p_e=(E_e,\mathbf{p}_e)$ is the positron momentum, $q=(\nu,\mathbf{q})$ is the virtual gauge boson loop momentum, $x=\hat{\mathbf{p}}_e\cdot\hat{\mathbf{q}}$ is the angle between the positron and emitted boson, and $\beta=|\mathbf{p}_e|/E_e$ the positron velocity. The weak-boson propagator has been left in the integrand for completeness, although the relevant nuclear response is concentrated at momentum scales far below $M_W$. All nuclear structure information is contained in the spin-invariant, parity-odd amplitude $ T_{3} $ that appears in Eq.~\eqref{section:superallowed:equation:gW_box}. For spin-zero external states, this is the only surviving piece of the Compton tensor and therefore the only relevant part of the hadronic tensor for the $ \gamma W $ box correction; a complete expression in terms of the multipole expanded currents is provided in the following section. It is conventionally defined as~\cite{PRC.107.035503.2023}
\begin{equation}
    T^{\mu \nu} \big\vert_{ \mathrm{VA} }
    = \frac{ i \epsilon^{ \mu \nu \alpha \beta } p_{\alpha} q_{\beta} }{ 2 p \cdot q } \ T_{3}
    \quad ,
\label{section:superallowed:equation:T3_decomposition}
\end{equation}
where the factor of $ 2 p \cdot q $ is convention and is not to be confused with the normalization of the tensor itself. We note the convention $ \epsilon^{0123} = -1 $ in the above expression.

We now comment on the contact term $ B_{\mathrm{em}\,\mathrm{W}}^{\mu\nu}(\mathbf{q}) $ in the non-relativistic hadronic tensor. This term arises from physical states at scales not described by the low-energy effective theory, e.g., multi-hadron intermediate states or nuclear shadowing effects, and these states are therefore not explicitly present in the spectral decomposition of the hadronic tensor in Eq.~\eqref{section:hadronictensor:equation:emw_explicit}. Now, we are only interested in the anti-symmetric part of the Compton tensor $ T_{3} $ and so, when contracted with the Levi-Civita in Eq.~\eqref{section:superallowed:equation:T3_decomposition}, the term in the photon propagator with the gauge fixing parameter which goes as $ \sim (1-\xi) \, q_{\mu} q_{\nu} $ vanishes. Unlike in two-photon exchange where a seagull vertex is present to guarantee gauge invariance of the hadronic tensor (represented in Coulomb gauge), the gauge fixing parameter is irrelevant and the result is manifestly gauge invariant. The contributions which appear in $ B_{\mathrm{em}\,\mathrm{W}}^{\mu\nu}(\mathbf{q}) $ are not needed to regulate any divergences but are associated to physics with quark-gluon degrees of freedom, for example, the free-neutron $ \gamma W $ box function.

While $ T^{\mu \nu} $ is built from relativistically normalized states, one needs to match onto the amplitudes in non-relativistic nuclear many-body theory, i.e., to match onto the unit-normalized tensor in Eq.~\eqref{section:hadronictensor:equation:emw_explicit}. At leading order in the heavy-particle limit, the amplitude-level matching between the two descriptions amounts to an overall normalization of the external states as
\begin{align}\label{section:hadronictensor:equation:nonrelativistic_amplitude}
    \bar{u}_{f}(\mathbf{p}_{f}) \, \Gamma_{O} \, u_{i}(\mathbf{p}_{i})
    = \sqrt{ 4 M_f M_i } \, \langle f | \int d^{3}x \ O(x) \ | i \rangle + \cdots
    \quad ,
\end{align}
where $ \mathcal{O}(x) $ is a given current operator with Dirac matrix representation $ \Gamma_{O} $, and the ellipses denote relativistic corrections of $ \mathcal{O}(\langle \mathbf{p}^2 \rangle / M ) $ to the amplitude (see Section XII–1 of Ref.~\cite{CUP.2.2014} for a comprehensive discussion). Applying Eq.~\eqref{section:hadronictensor:equation:nonrelativistic_amplitude} to the Compton tensor gives $ T^{\mu \nu} = i/2 \sqrt{ 4 M_{f} M_{i} } \ H^{\mu \nu}_{ \mathrm{em} \, \mathrm{W} } + \cdots $ where the normalization of the intermediate spectrum cancels against the relativistic completeness measure. The relativistic amplitude $ T_{3} $ entering the $ \gamma W $ box can therefore be evaluated entirely with unit-normalized eigenstates of the non-relativistic nuclear Hamiltonian.

Thus, the task is to compute $T_3$ with sufficient nuclear-structure input and to insert it into the loop integral defining the $\gamma W$ box function, after which one averages over the positron phase space to obtain
\begin{align}
    \square_{ \gamma W }^{b, \mathrm{nuc.}} = \big\langle \, \mathrm{Re} \, \square_{ \gamma W }^{b, \mathrm{nuc.}} (E_{e}) \, \big\rangle
    = \frac{ \int_{ m_{e} }^{ E_{0} } dE_{e} \ \vert \vec{p}_{e} \vert \, E_{e} ( E_{0} - E_{e} )^{2} \, F( Z_{f}, E_{e} ) \,
    \mathrm{Re} \, \square_{ \gamma W }^{b, \mathrm{nuc.}} (E_{e}) }
    { \int_{ m_{e} }^{ E_{0} } dE_{e} \ \vert \vec{p}_{e} \vert \, E_{e} ( E_{0} - E_{e} )^{2} \, F( Z_{f}, E_{e} ) }
    \quad .
\end{align}
When properly combined with the state-of-the-art result for free neutron decay~\cite{PRL.121.241804.2018}, this provides the nuclear-structure-dependent electroweak radiative correction
\begin{equation}
    \delta_{\mathrm{NS}} = 2 ( \square_{\gamma W}^{b, \text{nuc.}} - \square_{\gamma W}^{n} )
    \quad ,
\end{equation}
to a given Fermi transition.


\subsection{The Forward Hadronic Tensor and the \texorpdfstring{$ \gamma W $}{gamma-W} Box Loop Integral}\label{section:superallowed:section:gW_box}

The nuclear amplitude $T_3$ is evaluated by expanding the electroweak currents in Coulomb, longitudinal, transverse electric, and transverse magnetic multipoles (see the discussion in Section~\ref{sec:multipoles_current_operators}). The expanded form is given by
\begin{align}
    T_3(\nu,\mathbf{q})
    &=
    4\pi i\,\frac{\nu}{|\mathbf{q}|}\sqrt{ M_i M_f} \sum_{J=1}^{\infty}(2J+1)
    \\
    & \mkern-36mu \mkern-9mu \times \bigg[ \; \langle f | \; \Big\{ T_{\mathrm{em},J0}^{\mathrm{el}} (z_f-H)^{-1} T_{\mathrm{W},J0}^{5,\mathrm{mag}}
    - iT_{\mathrm{em},J0}^{\mathrm{mag}} (z_f-H)^{-1} iT_{\mathrm{W},J0}^{5,\mathrm{el}} \Big\} \;|i\rangle
    \nonumber \\
    & \mkern-36mu \ \ \
    +
    \langle f | \; \Big\{ T_{\mathrm{W},J0}^{5,\mathrm{mag}} (z_i-H)^{-1} T_{\mathrm{em},J0}^{\mathrm{el}} - iT_{\mathrm{W},J0}^{5,\mathrm{el}} (z_i-H)^{-1} iT_{\mathrm{em},J0}^{\mathrm{mag}} \Big\} \; |i\rangle \; \bigg]
    \quad ,
    \label{eq:review-t3-multipole}
\end{align}
where the label ``$ 5 $'' denotes a multipole arising from the axial-vector part of the weak-charged current, the arguments of the resolvents contain $ z_{i} = M_{i} - \nu + i \varepsilon_{3} $ and $ z_{f} = M_{f} + \nu + i \varepsilon_{3} $, and we have further made use of Eq.~\eqref{section:hadronictensor:equation:nonrelativistic_amplitude} to convert the amplitude into its non-relativistic form. The amplitudes of the resolvent are handled in full by use of the Lanczos methods described in Section~\ref{Section: 3.2}. In pragmatic calculations, the sum over multipoles is truncated at some finite $ J $ to reach the desired precision.

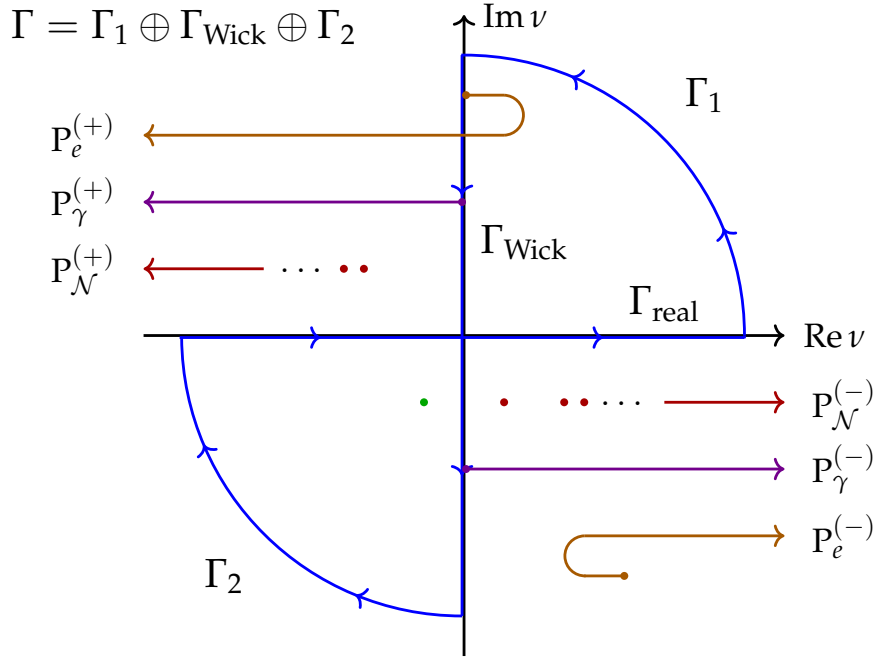
\begin{figure}[t!]
\center\resizebox{0.75\textwidth}{!}{%
\begin{tikzpicture}
[
	decoration={%
	markings,
	mark=at position 0.25 with {\arrow[line width=1pt]{>}},
	mark=at position 0.75 with {\arrow[line width=1pt]{>}},
	}
]
\def\deltaaxis{0.025}
\def\contourradius{3.5}
\def\deltapolesnuclear{0.833}
\def\deltapolesphoton{1.666}
\def\deltapoleselectron{2.5}
\draw [ ->, line width=1.0 ] (-4,0) -- (4,0) node[ right, font=\large ] {$ \ \real \nu $};
\draw [ ->, line width=1.0 ] (0,-4) -- (0,4) node[ right, font=\large ] {$ \ \imaginary \nu $};
\node [ font=\Large ] at ( -\contourradius, \contourradius*1.1 ) {$ \Gamma = \Gamma_{1} \oplus \Gamma_{ \mathrm{Wick} } \oplus \Gamma_{2} $};
\path [ draw, blue, line width=1.0, postaction=decorate ] ( \contourradius, 0 ) arc ( 0 : 90 : \contourradius );
\path [ draw, blue, line width=1.0 ] ( 0, \contourradius ) -- ( -\deltaaxis - 0.01, \contourradius );
\node [ font=\Large ] at ( \contourradius - 0.5, \contourradius - 0.5 ) {$ \Gamma_{1} $};
\path [ draw, blue, line width=1.0, postaction=decorate ] ( -\deltaaxis, -\contourradius ) arc ( 90 : 0.5 : -\contourradius );
\node [ font=\Large ] at ( -\contourradius + 0.5, -\contourradius + 0.5 ) {$ \Gamma_{2} $};
\path [ draw, blue, line width=1.0, postaction=decorate ] ( -\contourradius - 0.05, -\deltaaxis ) -- ( \contourradius, -\deltaaxis );
\path [ draw, blue, line width=1.0 ] ( \contourradius, -\deltaaxis - 0.01) -- ( \contourradius, 0.0 );
\node [ font=\Large ] at ( \contourradius - 1.0, 0.4 ) {$ \Gamma_{ \mathrm{real} } $};
\path [ draw, blue, line width=1.0, postaction=decorate ]  ( -\deltaaxis, \contourradius ) -- ( -\deltaaxis, -\contourradius );
\node [ font=\Large ] at ( 0.75, \contourradius/3 ) {$ \Gamma_{ \mathrm{Wick} } $};
\draw[ fill=green!65!black ] ( -0.5, -\deltapolesnuclear ) coordinate [ circle, fill, inner sep=1pt ] (np0);
\draw[ fill=red!65!black ]
( -1.25, \deltapolesnuclear ) coordinate [ circle, fill, inner sep=1pt ] (np2)
( -1.5, \deltapolesnuclear ) coordinate [ circle, fill, inner sep=1pt ] (np3);
\node [ font=\large ] at ( -2.0, \deltapolesnuclear - \deltaaxis ) {$\cdots$};
\draw[ ->, line width=1.0, red!65!black ] ( -2.5, \deltapolesnuclear ) -- ( -4, \deltapolesnuclear );
\node [ font=\large ] at ( -4.75, \deltapolesnuclear ) {$ \mathrm{P}^{(+)}_{ \mathcal{N} } $};
\draw[ fill=red!65!black ]
( 0.5, -\deltapolesnuclear ) coordinate [ circle, fill, inner sep=1pt ] (np2)
( 1.25, -\deltapolesnuclear ) coordinate [ circle, fill, inner sep=1pt ] (np3)
( 1.5, -\deltapolesnuclear ) coordinate [ circle, fill, inner sep=1pt ] (np4);
\node [ font=\large ] at ( 2.0, -\deltapolesnuclear - \deltaaxis ) {$\cdots$};
\draw[ ->, line width=1.0, red!65!black ] ( 2.5, -\deltapolesnuclear ) -- ( 4, -\deltapolesnuclear );
\node [ font=\large ] at ( 4.75, -\deltapolesnuclear - \deltaaxis ) {$ \mathrm{P}^{(-)}_{ \mathcal{N} } $};
\draw [ fill=red!45!blue ] ( \deltaaxis, -\deltapolesphoton ) coordinate [ circle, fill, inner sep=1pt ] (pp2);
\draw [ ->, line width=1.0, red!45!blue ] ( 0, \deltapolesphoton ) -- ( -4, \deltapolesphoton );
\node [ font=\large ] at ( -4.75, \deltapolesphoton ) {$ \mathrm{P}^{(+)}_{ \gamma } $};
\draw [ fill=red!45!blue ] ( -\deltaaxis, \deltapolesphoton ) coordinate [ circle, fill, inner sep=1pt ] (pp1);
\draw [ ->, line width=1.0, red!45!blue ] ( 0, -\deltapolesphoton ) -- ( 4, -\deltapolesphoton );
\node [ font=\large ] at ( 4.75, -\deltapolesphoton ) {$ \mathrm{P}^{(-)}_{ \gamma } $};
\draw [ fill=red!65!green ] ( \deltaaxis, \deltapoleselectron + 0.50 ) coordinate [ circle, fill, inner sep=1pt ] (pp1);
\draw [ red!65!green, line width=1.0 ] ( 0.5, \deltapoleselectron ) arc ( 92 : 268 : -0.25);
\draw [ red!65!green, line width=1.0 ] ( 0.0, \deltapoleselectron + 0.50 ) -- ( 0.5, \deltapoleselectron  + 0.50 );
\draw [ ->,  red!65!green, line width=1.0 ] ( 0.5, \deltapoleselectron ) -- ( -4.0, \deltapoleselectron );
\node [ font=\large ] at ( -4.75, \deltapoleselectron ) {$ \mathrm{P}^{(+)}_{ e } $};
\draw [ fill=red!65!green ] ( 2.0, -\deltapoleselectron - 0.50 ) coordinate [ circle, fill, inner sep=1pt ] (pp2);
\draw [ red!65!green, line width=1.0 ] ( 1.5, -\deltapoleselectron ) arc ( 92 : 272 : 0.25);
\draw [ red!65!green, line width=1.0 ] ( 2.0, -\deltapoleselectron - 0.50 ) -- ( 1.5, -\deltapoleselectron  - 0.50 );
\draw [ ->,  red!65!green, line width=1.0 ] ( 1.5, -\deltapoleselectron ) -- ( 4.0, -\deltapoleselectron );
\node [ font=\large ] at ( 4.75, -\deltapoleselectron ) {$ \mathrm{P}^{(-)}_{ e } $};
\end{tikzpicture}%
}%
\captionsetup{justification=raggedright}
\caption{\label{section:gW_box:figure:contour_deformation}
Example trajectories of the $\nu$-integral poles in Eq.~(\ref{section:superallowed:equation:gW_box}) arising from the dynamics of: (i) the nuclear propagators in $ T_{3} $, plotted in red for $ \Delta_{n} > 0 $ and green for $ \Delta_{n} < 0 $; (ii) the photon propagator, plotted in purple; and (iii) the electron propagator, plotted in orange. The corresponding sets are labelled by $ \mathcal{N} $, $ \gamma $ and $ e $, respectively. The chosen contour deformation $ \Gamma $ is the oriented contour defined by $ \Gamma = \Gamma_{1} \oplus \Gamma_{ \mathrm{Wick} } \oplus \Gamma_{2} $, shown in blue.
}
\end{figure}

The loop integral in Eq.~\eqref{section:superallowed:equation:gW_box} contains poles from the photon propagator, the electron propagator, and the nuclear propagators in $ T_{3} $, so direct principal value integration over real $\nu$ is inconvenient. As in Ref.~\cite{PRL.134.012501.2025}, one may deform the $\nu$ contour to the imaginary axis, $\nu=i\nu_E$, while accounting for any poles enclosed by the contour deformation. Tracking the various poles in Figure~\ref{section:gW_box:figure:contour_deformation} then gives
\begin{equation}\label{section:amplitudes:equation:contour_deformation}
    \square_{\gamma W}^{b}
    =
    \left(\square_{\gamma W}^{b}\right)_{\mathrm{Wick}}^{\uparrow}
    +
    \left(\square_{\gamma W}^{b}\right)_{\mathrm{Res},T_3}
    +
    \left(\square_{\gamma W}^{b}\right)_{\mathrm{Res},e}
    \quad .
\end{equation}
The first term is the Wick-rotated integral with the direction of integration upwards along the $ \mathrm{Im} \, \nu $ axis. The second term collects residues of the Compton amplitude located in quadrant III which arise from low-lying states in the intermediate spectrum below the final state $ 0^{+} $~\cite{PRL.134.012501.2025}, with
\begin{align}\label{section:superallowed:equation:compton_poles}
    \nu_n = \Delta_n + \frac{|\mathbf{q}|^2}{2M_n} - i\varepsilon
    \qquad \quad
    \Delta_n = M_n - M_f < 0
    \quad . \nonumber
\end{align}
The Compton residue contribution is incurred only on the momentum interval
\begin{equation}\label{section:superallowed:equation:compton_domain}
    \vert \mathbf{q} \vert \le \sqrt{ -2 \Delta_{n} M_{n} }
    \quad ,
\end{equation}
after which the pole escapes the contour. The third term is generated by the electron propagator poles enclosed by the contour in quadrant I; we neglect to discuss this contribution further since it has been shown to be negligible by several orders of magnitude. Contributions from the large semicircles vanish in a given non-relativistic nuclear model, and the pole associated with the $W$ propagator lies at momenta of order $M_W$, well outside the support of a non-relativistic calculation.

The Wick term has the generic form
\begin{equation}
    \left(\square_{\gamma W}^{b}\right)_{\mathrm{Wick}}(E_e)
    = e^2 \int_0^\infty d|\mathbf{q}| \int_{-\infty}^{\infty} d\nu_E \ K ( i\nu_E,|\mathbf{q}|;E_e) \ \frac{i \, T_3(i\nu_E,\mathbf{q})}{M f_+},
\end{equation}
and the Compton residue term
\begin{equation}
    \left(\square_{\gamma W}^{b}\right)_{\mathrm{Res,T3}}(E_e)
    = e^2 \int_0^\infty d|\mathbf{q}| \ K ( \nu_k,|\mathbf{q}|;E_e) \ \frac{i \, \mathrm{Res} \, T_3(\nu_k,\mathbf{q})}{M f_+},
\end{equation}
where the function $ K $ contains the photon, electron, and $ W $-boson propagators with the angular integration in Eq.~\eqref{section:superallowed:equation:gW_box} having been performed analytically. The appearance of $ K $ in the Wick contribution is understood as the analytic continuation of the function $ K $ as it would appear in Eq.~\eqref{section:superallowed:equation:gW_box}. The decomposition in Eq.~\eqref{section:amplitudes:equation:contour_deformation} then provides a controlled way of computing $ \square_{\gamma W}^{b}(E_{e}) $ on the real-$\nu$ axis.


\subsection{\textit{Ab-initio} Calculations in Light Nuclei and Impact on CKM Unitarity}\label{section:superallowed:section:abinitio}

In Table~\ref{section:superallowed:table:results}, we summarize the state-of-the-art results for the electroweak radiative corrections to the $ \carbonfermidecay $ transition with $ \delta_{\mathrm{NS}} $ as computed in the current algebra (CA) formalism, along with its closest analog in the EFT formalism, $ \delta_{\mathrm{NS}}^{(0)} |_{\mathrm{mag+LS}} $. We note that while there is significant overlap in the physics which enters these two quantities, they are not directly comparable and there is an ongoing effort to perform a rigorous matching between the two formalisms (private communication from C.-Y. Seng). Most important to observe is that the modern results are discrepant with the values adopted in the Hardy and Towner 2014 critical survey on superallowed decays~\cite{PRC.91.025501.2015}. The observed shift confirms that an \textit{ab-initio} treatment of the nuclear response with genuine many-body approaches gives a numerical enhancement to prior estimates of the same physics.

\begin{table}[t!]
    \centering
    \caption{\label{section:superallowed:table:results}
    A summary of state-of-the-art results for evaluations of $ \delta_{\mathrm{NS}} $ in the current algebra (CA) formalism and the long-range part of the radiative correction $ \delta_{\text{NS}}^{(0)} |_{\mathrm{mag+LS}} $ in the EFT formalism for the $ \carbonfermidecay $, as compared to the last two available surveys on superallowed decays from Hardy and Towner. Results are quoted in the table as provided in their corresponding references.}
    \def\arraystretch{1.25}
    \begin{tabular}{ll}
    \toprule
        $ [ \% ] $ & $ \carbonfermidecay $ \\
    \midrule
        CA $ \delta_{\mathrm{NS}} $ & $ -0.422 (29)_{\mathrm{nuc}} (12)_{n, \mathrm{el}} $~\cite{PRL.134.012501.2025} \\
        EFT $ \delta_{\text{NS}}^{(0)} |_{\mathrm{mag+LS}} $ & $ - [ \, 0.406, \, 0.443 \, ] $~\cite{PRL.133.211801.2024} \\
        H\&T 2020~\cite{PRC.102.045501.2020} & $ -0.400(35)(36) $ \\ 
        H\&T 2014~\cite{PRC.91.025501.2015} & $ -0.345(35) $ \\ 
    \bottomrule
    \end{tabular}
\end{table}

As detailed in Ref.~\cite{PRL.134.012501.2025} for the $ \carbonfermidecay $ transition, the central value for $ \delta_{\mathrm{NS}} $ in the CA formalism is obtained by averaging over two realistic, different-order chiral interactions at fixed harmonic oscillator frequency, while the uncertainty combines the many-body truncation, multipole truncation, one-nucleon form-factor dependence, subtraction of the elastic single-nucleon $ \gamma W $ box, chiral truncation, residual oscillator-frequency dependence, and the so-called nuclear-shadowing uncertainty~\cite{PRD.76.094023.2007, PR.512.255393.2012, PPNP.68.314372.2013, JPG.32.R367.2006}. The prediction from the more rigorous calculation agrees with the most recently adopted value for the $ \carbonfermidecay $ decay in the Hardy and Towner 2020 critical survey of superallowed decays~\cite{PRC.102.045501.2020}; the uncertainty is further reduced by a factor of $ 1.6\mathrm{x} $ back to its estimated 2014 value, despite the inclusion of the largely unconstrained nuclear shadowing effects in the uncertainty budget. Unfortunately, Bayesian analyses are presently prohibitive.

The complementary treatment recently developed in Refs.~\cite{arXiv:2605.14006, PRL.133.211801.2024, PRC.110.055502.2024} sees the nucleus-dependent radiative corrections relevant for $ \mathcal{O}(10^{-4}) $ precision derived from a tower of EFTs running from the electroweak scale to hadronic and nuclear scales. In contrast to the current algebra approach with explicit responses of the intermediate spectrum, the EFT approach separates hard, soft, potential, and ultrasoft photon modes according to the hierarchy of scales $ \vert \mathbf{q}_{\mathrm{ext}} \vert \sim m_{e} \ll m_{\pi} \ll \Lambda_{\chi} \ll M_{W} $. Many formal developments of this approach began with the treatment of electromagnetic bound states in potential non-relativistic quantum electrodynamics~\cite{PRD.58.093013.1998, PRD.59.016005.1998, CJP.77.267278.1999, PRL.85.22482251.2000}. This allows for a transparent factorization of the decay rate at the desired precision. The short-range contribution is governed by two presently unknown low-energy constants, denoted $g^{NN}_{V1}$ and $g^{NN}_{V2}$, so the framework is systematic but not parameter free. These constants may be obtained either from matching to QCD or, possibly, from a fit to several precisely measured superallowed transitions. With this approach in a given many-body method, one merely needs to evaluate two-body amplitudes between the relevant initial and final states to predict the complete radiative correction $ \bar{\delta}_{\mathrm{NS}}^{E} $. We include results for the $ \carbonfermidecay $ transition as taken from the second of two independent studies in this formalism using Quantum Monte Carlo (QMC) approaches; the study on the $ \oxygenfermidecay $ decay is not readily comparable to the historical values. The quoted value contains only the long-range part of the correction since this approximately maps onto the current algebra formalism. The overall uncertainty in $ \bar{\delta}_{\mathrm{NS}}^{E} $ is dominated by the unknown short-range couplings which a rigorous fit could reduce to $ \delta V_{ud} \sim 3\times10^{-4} $, competitive with that of the current algebra calculation.

The impact on $ V_{ud} $ is therefore twofold. First, the current algebra calculation reduces the theory uncertainty assigned to the lightest superallowed Fermi transition, where the radiative correction is the central limitation on the corrected $ \mathcal{F}t $ value. Second, the shift of all modern calculations relative to historical evaluations changes the inferred degree of consistency among superallowed emitters and hence the weight carried by $ \carbonfermidecay $ and $ \oxygenfermidecay $ in tests of CKM unitarity. This is doubly important because the light decays $ \carbonfermidecay $ and $ \oxygenfermidecay $ are by far the most sensitive to possible scalar-current extensions of the SM~\cite{PRC.102.045501.2020}. The promise of further improvements to theory extractions of radiative corrections also clarifies the target for experiment. For $ {}^{10}\mathrm{C} $, the dominant experimental limitation remains the branching ratio. Coupled with the reduction in radiative correction uncertainties, a new high-precision branching-ratio measurement would therefore feed into (i) improving the contribution of the $ \carbonfermidecay $ to a more competitive determination of $ V_{ud} $, and (ii) enforcing increasingly strict constraints on scalar extensions of the SM. Proposed recoil-detection approaches, including superconducting tunnel junction measurements inspired by successful $ {}^{7}\mathrm{Be} $ electron-capture studies~\cite{PRL.125.032701.2020, JLTP.209.796803.2022, PRXQ.4.010315.2023}, and trapped-ion recoil programs sensitive to exotic $\beta$-decay currents~\cite{PRL.110.092502.2013, PRL.128.202503.2022}, are already occurring in this direction. In the next decade, improved theory and targeted measurements become tightly coupled ingredients in extracting $ V_{ud} $ with higher and higher precision in motion toward the ultimate goal of resolving the Cabibbo anomaly.


\section{Conclusions and Future Directions}\label{sec:conclusions_future_directions}

Today, several precision tests of the Standard Model at the low-energy frontier are limited by nuclear-structure theory and its associated uncertainties. In this review, we have discussed the leading nuclear-structure corrections entering two such programs -- the two-photon-exchange contribution to the muonic Lamb shift and the $\gamma W$ box radiative correction to superallowed decays -- and shown that they are two faces of a single theoretical coin: the generalized hadronic tensor of Section~\ref{sec:common_theory}. The modern \textit{ab-initio} approach, rooted in $\chi$EFT Hamiltonians and consistent electroweak currents along with Bayesian uncertainty quantification, has turned the evaluation of nuclear electroweak correlators from model-dependent estimates into systematically improvable calculations.

The dividends of such progress are visible in the milestones reviewed herein. In muonic atoms (Section~\ref{sec:muonic_atoms}), the N$^3$LO computation of the two-photon-exchange corrections in muonic-helium ions, equipped with Bayesian truncation errors, enabled precise determinations of the helion and $\alpha$-particle charge radii~\cite{Ji18,Kra21,Sch25,Lim25}. For the helium isotope-shift puzzle, these rigorously quantified uncertainties clearly excluded nuclear structure corrections as the origin of the $4\sigma$ tension and directed the scrutiny toward atomic theory, where the missing contributions were ultimately found~\cite{Qi25,Pac24}. This puzzle was thus resolved without the need to invoke new physics. In superallowed $\beta$ decays (Section~\ref{section:superallowed}), a first-of-its-kind \textit{ab-initio} computation of the nuclear $\gamma W$ box for the lightest transition, $\carbonfermidecay$~\cite{PRL.134.012501.2025}, re-sharpened the evaluation of the nuclear-structure-dependent correction that dominates the $V_{ud}$ uncertainty budget~\cite{PRC.102.045501.2020,ARNPS.74.2347.2024} and established a systematically improvable approach towards confronting the $2$--$3\sigma$ tension in top-row CKM unitarity test~\cite{Nav24, ARNPS.74.2347.2024}.

The road ahead still presents several challenges. On the side of methodology, fully uncertainty-quantified predictions require propagating the statistical uncertainties of the low-energy constants of $\chi$EFT into electroweak observables. At the moment, this is only feasible with reduced-order emulators such as eigenvector continuation (Section~\ref{sec:uncertainty_quantification}). Greater precision further requires the inclusion of electroweak currents at higher chiral orders via consistent expansion of both electroweak currents and nuclear Hamiltonians in $\chi$EFT, as well as computational advances to render larger model spaces accessible. On the physics side, experiment sets the agenda. The QUARTET and muX programs~\cite{Oha24,Ung24,Wau21,Ada23} are working to extend muonic-atom spectroscopy beyond the light-mass region, where nuclear polarization remains the limiting theoretical input and bridging exact \textit{ab-initio} methods with the mean-field-based approaches of heavier systems~\cite{Rat25,Val22} is a significant challenge. Another important observable pursued by the CREMA collaboration is the hyperfine splitting~\cite{Nub22}, which likewise requires nuclear-structure input, but of a spin-dependent nature that is considerably more challenging to compute~\cite{Fri05,Zha24,Bon26}. For superallowed decays, the completion of the $\oxygenfermidecay$ analysis and the extension to heavier emitters such as ${}^{18}\mathrm{Ne}$, ${}^{22}\mathrm{Mg}$, and ${}^{26m}\mathrm{Al}$, together with \textit{ab-initio} evaluations of the isospin-symmetry-breaking correction $\delta_{\mathrm{C}}$, will determine whether the unitarity deficit survives such improvements from theory. Indeed, the two programs reviewed here already intersect in this arena: the charge radii entering the ${}^{26m}\mathrm{Al}$ analysis are themselves obtained from muonic X-ray spectroscopy, and an updated nuclear polarization correction has recently been shown to reduce the CKM unitarity deficit by one standard deviation~\cite{PRR.7.L042002.2025}.

Notably, the same machinery discussed in this review readily generalizes to (i) the $\gamma Z$ box, which enters weak-charge measurements, (ii) to muon capture, a process with strong connections to neutrinoless double-$\beta$ decay, and (iii) to neutrino-nucleus scattering. With percent-level, statistically meaningful control over electroweak processes in nuclei within reach, low-energy precision experiments stand as competitive and complementary probes to direct searches at colliders, sensitive to new physics at scales of $10$~TeV and beyond.


\authorcontributions{The authors have contributed equally in the preparation of this review.}

\funding{This work was supported by the Swedish Research Council via a Research Project Grant, No.\ 2021-04507. This work was supported in part by the Deutsche Forschungsgemeinschaft (DFG) through the Cluster of Excellence ``Precision Physics, Fundamental Interactions, and Structure of Matter'' (Project ID 390831469).}


\dataavailability{No new data were created or analyzed in this study. Data sharing is not applicable to this article.}

\acknowledgments{We thank Chien-Yeah Seng, Wouter Dekens, Petr Navvratil, Ben Ohayon, Franziska Hagelstein and Vadim Lensky for helpful comments on the manuscript.}

\conflictsofinterest{The authors declare no conflicts of interest.}

\reftitle{References}

\bibliography{
    ./bibfiles/simone_bibfile,
    ./bibfiles/lanczos,
    ./bibfiles/nucth,
    ./bibfiles/gW_box
}

\end{document}